# Nano-engineered surface enhanced Raman spectroscopy substrates for probing tissue-material interactions

*Connie M. Wang[1]*, Roberta M. Sabino[2], Aditya Garg[3], Ahmed E. Salih[3], Loza F. Tadesse[3,6,7]*, Elazer R. Edelman[4,5]*

[1]Department of Biological Engineering, MIT, Cambridge, MA 02139, USA

[2]Department of Chemical and Biomedical Engineering, University of Wyoming, Laramie, WY 82071, USA

[3]Department of Mechanical Engineering, MIT, Cambridge, MA 02139, USA

[4]Institute for Medical Engineering and Science, MIT, Cambridge, MA 02139, USA

[5]Cardiovascular Division, Brigham and Women's Hospital, Harvard Medical School, Boston, MA, 02115, USA.

[6]Ragon Institute of MGH, MIT and Harvard, Cambridge, MA 02139, USA

[7]Jameel Clinic for AI & Healthcare, MIT, Cambridge, MA 02139, USA

* **Corresponding author (email):** Connie M. Wang (cwang1999@gmail.com), Loza F. Tadesse (lozat@mit.edu)

## ABSTRACT

Innovation in biomaterials has brought both breakthroughs and new challenges in medicine, as implant materials have become increasingly multifunctional and complex. One of the greatest issues is the difficulty in assessing the temporal and multidimensional dynamics of tissue-implant interactions. Implant biology remains hard to decipher without a noninvasive and multiplexed technique that can accurately monitor real-time biological processes. To address this, we developed a multifunctional, self-sensing implant material composed of gold nano-columns patterned on a titanium surface (AuNC-Ti). This material acts as a nanoengineered surface-enhanced Raman spectroscopy (SERS) substrate that amplifies biological Raman signals at the tissue-implant interface, providing the ability to sense tissue-material interactions in a multiplexed and nondestructive manner. AuNC-Ti SERS substrates were fabricated using oblique angle deposition (OAD) and characterized using scanning electron microscopy (SEM) to show uniform formation of AuNCs (360 ± 40 nm in length and 50 ± 16 nm in width). X-ray photoelectron spectroscopy (XPS), X-ray diffraction (XRD), and contact angle measurements demonstrated biocompatible surface chemistry with ideal wettability. Biocompatibility was further demonstrated via *in vitro* cytotoxicity assays on human aortic endothelial cells (HAECs) cultured on AuNC-Ti surfaces. The median SERS enhancement factor (EF) was calculated to be $1.8 \times 10^5$, and spatial identification of reporter molecules and porcine tissue components on AuNC-Ti surfaces was demonstrated using confocal Raman imaging and multivariate analysis. Our approach utilizes unlabeled SERS and machine learning, promising multiplexed characterization of tissue-material interactions and subsequently enabling tissue state determination and non-invasive monitoring of implant-tissue interaction.

## INTRODUCTION

Biomaterials have long surpassed their original role as biologically inert implants to now assume a more central and active function in therapeutics. Drug-eluting polymeric materials, surface-modified metals, and bio-mimetic scaffolds are now integral to modern medicine with tremendous impact on patient health.[1] Yet, the increasing sophistication of these implants is also accompanied by failure from the equally complex tissue responses they elicit. Indeed, we have learned much from biomaterial science as to the pace and nature of tissue repair and the foreign body response, and the centrality of chronic inflammation, fibrotic encapsulation, and infection.[2] There is not though the same degree of sophistication in detecting these events – they thus proceed and escalate long before they are recognized. Current monitoring of implant healing relies on biomarker detection through laboratory blood tests, live imaging such as magnetic resonance imaging (MRI) and X-ray, histology, and more recently, omics analysis.[3–5] Circulating markers provide an idea of systemic activation but are only acquired intermittently and reported after a lag from time of sampling. High resolution imaging can localize identification with spatial precision but cannot be performed repeatedly, and histology and omics are destructive and cannot provide data in real time. In short, there is not a means of attaining high fidelity, real-time simultaneous assessment of the temporal and multidimensional nature of biological systems in response to foreign materials *in vivo*.

Surface-enhanced Raman spectroscopy (SERS) might address this gap by enabling noninvasive yet multiplexed sensing of complex biological tissue at an implant material surface. Raman spectroscopy relies on detecting the inelastic scattering of light, which can be translated to a unique spectrum for chemical characterization. Though Raman scattering is a weak process occurring in ~one in 10 million scattered photons, these signals can be amplified up to six orders of magnitude for molecules adsorbed on noble metal nanostructures via plasmonic interaction.[6] This process is termed surface enhanced Raman spectroscopy (SERS).[7] SERS signals can be detected at depths between 100-200 μm and up to 50 mm *in vivo* when collected using spatially offset Raman spectroscopy (SORS).[8] SERS nano-probes have been proposed for labeled *in vivo* cancer detection and unlabeled detection of bioanalytes *in vitro*.[7,9,10] Labeled SERS has found success in deep tissue, *in vivo* imaging applications but with limited multiplexity. Unlabeled SERS on the other hand can encompass the chemically complex nature of biological processes and capture high-dimensional spectral data from native molecules rather than probes. Surfaces patterned with gold or silver nanoparticles known as SERS substrates have been engineered to facilitate unlabeled detection analytes placed directly on the surface, requiring minimal sample preparation. In combination with multivariate analysis and deep learning models, unlabeled SERS has enabled powerful sensing capabilities including bacteria identification, virus detection, and disease state classification.[11–13]. Unlabeled spatially offset SERS (SESORS) has also found success in deep tissue detection of neurotransmitters in concentrations as low as 100 μM using colloidal gold nanoparticles.[14] Such advancements indicate the potential of unlabeled SESORS and implantable SERS-substrates as a self-sensing biomaterial to aid us in understanding the interplay between material and tissue.

To our knowledge, there has yet to be an application of unlabeled SERS with demonstrated multiplexed sensing of implant fate, but there are multiple promising examples of labeled SERS-active implants. These include work by Yang et al. and their development of an implantable AuNP and Si-based substrate for glucose sensing in eyes for diabetes management.[15] Label-based AuNP-functionalized polypropylene meshes have also been explored by Lanzalaco et al. for detecting inflammation in post-operative patients via thermal sensing.[16] These SERS technologies have made remarkable progress toward developing implantable sensors for minimally-invasive treatment of disease. However, a process complex and dynamic as peri-implant healing of dental or orthopedic implants would likely demand higher resolution for sensing stages such as initial protein surface adsorption, matrix remodeling, cell migration, and finally de novo bone formation.[17] Dental and orthopedic implants also require highly specific implant material properties differing from previous SERS substrate bases such as silicon chips or polymer meshes.[18] This leaves us with an exciting opportunity to extend previous innovations in implantable SERS devices to additional clinically relevant materials in orthopedics, enabling us to investigate new SERS substrate materials for unlabeled sensing of tissue-implant interactions with newfound multiplexity and resolution.

Titanium is a prime implant material and possible SERS substrate when surface modified with noble metal nanostructures. Ti is one of the most common orthopedic and dental implant materials with tunable surface properties to adhere to other metals such as gold nanoparticles without sacrificing biocompatiblity[18,19]. The popularity of Ti in implants also stems from its mechanical strength, low density, corrosion resistance, and low modulus similar to that of bone which enhances osseointegrative processes.[20] Orthopedic and dental replacements are rising with the global market value projected to reach over USD 80 billion by 2030.[18,21,22] Revision surgeries are a growing concern with studies citing revision arthroplasty rates of 5% after five years and 12% after ten years.[23] The need for revision arises after complications such as implant loosening, infection, excessive wear, mechanical failure, modulus mismatch, and chronic inflammation which lead to implant failure – events that might not become so detrimental if detected early.[24] Revision surgeries are costly, painful, and result in double the rates of bacterial infection, lowering overall success.[24] Minimizing the need for revision arthroplasty is of high priority to clinicians especially as lifespan increases. Smart implants have been developed to reduce implant failure and include smart orthopedic and dental implants utilizing resonance frequency and tibial motion to monitor osseointegration, the healing process leading to stable fusion of bone and implant.[25,26] While these significant advancements have reached the clinic, informing surgeons of healing progress using non-invasive and comparatively inexpensive means, they are still limited by the amount and type of information they receive. Temperature and biochemical sensors have also been explored, but biochemical detection is currently limited to glucose[26] with mixed opinions on the effectiveness of such devices in practice[26]. There is then the potential for improvement in orthopedic and dental smart implant technology that we address through the application of SERS.

We present the development of gold nanocolumn-coated titanium (AuNC-Ti) as a SERS-active implant material. Oblique angle deposition (OAD) was chosen to fabricate the AuNC coating as a previously demonstrated, relatively simple, and reproducible method to create SERS substrates.[27] While OAD is typically used with a glass or silicon wafer substrate, it has yet to be demonstrated on an implant material like Ti to produce a SERS substrate for probing the tissue response to implant materials. We therefore fabricated and characterized a Ti-based SERS substrate composed of mechanically robust, biocompatible materials and validated its spatial Raman enhancing capabilities with principal component and vector component analysis (PCA and VCA). The AuNC-Ti material enhanced, detected, and spatially identified tissue components indicative of physiological state using Raman imaging and machine learning. This multifunctional material holds promise in advancing post-operative implant care through personalization of treatment plans, as clinicians will be able to understand and view the internal dynamics of peri-implant healing in real-time and thus act accordingly.

## RESULTS AND DISCUSSION

### *Fabrication of AuNC-Ti and surface morphology characterization*

We developed a SERS substrate coating composed of gold nanocolumns (AuNCs) on Ti, as an implantable material capable of enhancing relevant Raman signals at the implant-tissue interface to enable continuous local tissue environment monitoring. We coated Ti substrates with a nanocolumnar-like film of gold using oblique angle deposition (OAD) (**Figure 1a**). OAD relies on physical vapor deposition at a large angle (approaching 90°) to form a uniformly distributed coating of tilted nanocolumns that can produce effective, reproducible SERS substrates.[28] This method was chosen for its relatively simple process and previous success in *in vitro* studies such as malaria detection in blood cells and trace detection of methamphetamine.[13,29] Other methods for SERS substrate fabrication include synthesis of colloidal nanoparticles and subsequent coating on a substrate or photolithographic methods to create highly ordered plasmonic structures. These methods often require extensive use of cytotoxic chemicals and complex multi-step processes, thus for a clinically translatable implant application, there is a need for a biocompatible, low risk and reproducible synthesis procedure.[30] OAD was therefore chosen as a generally reliable and straightforward physical deposition method to produce uniform and nontoxic coatings with demonstrated SERS enhancement capabilities.[27]

Scanning electron microscopy (SEM) imaging **(Figure 1b)** confirmed a fairly uniform deposition of Au nanocolumns (AuNCs) with dimensions of 360 ± 40 nm in length, 50 ± 16 nm in width, and distribution density of 70 ± 10 AuNCs per $\mu m^2$ at an angle of 22 ± 4° from the polished Ti substrate surface. Respective measurement distributions are shown in **Figure 1c** and exhibited normal and lognormal distributions. The lognormal distribution of AuNC width is likely due to the occasional fusing of neighboring rods resulting in larger thicknesses. This slanted-rod morphology mirrors the structures in other OAD-fabricated SERS substrates in terms of nanocolumn dimensions and density, despite using an unconventional substrate material such as Ti.[27]

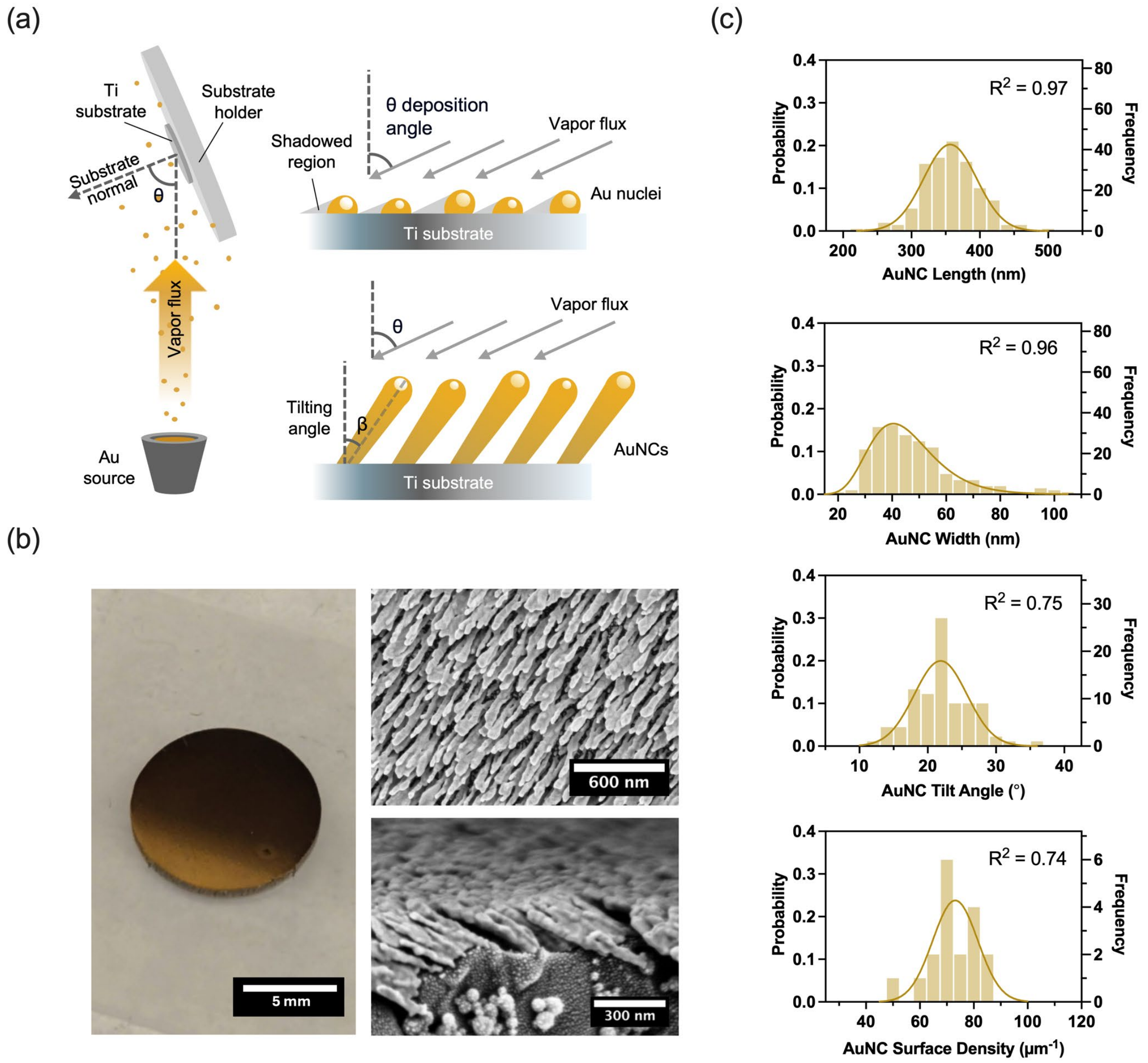

**Figure 1. a)** Schematic of AuNC-Ti fabrication using OAD. **b)** Image of whole AuNC-Ti sample and SEM image of AuNC-Ti surface with top view and side view. **c)** Probability and frequency distributions of AuNC dimensions: length, width, tilt angle, and surface density. All probability density functions follow a normal (AuNC length, tilt angle, and surface density) or lognormal (AuNC width) distribution.

***AuNC-Ti surface chemistry and crystal structure analysis***

Further characterization of the material surface before and after AuNC-Ti deposition was performed using X-ray photoelectron spectroscopy (XPS) and x-ray diffraction analysis (XRD) to identify any contaminants or superficial crystal phase changes. A balance must be struck in modifying implants for reporting the local milieu and eliciting or exacerbation the very negative reactions we seek to detect. No one single biological event drives tissue response to implanted materials. Implant fate hinges on the cumulative effects of a number of processes at the implant-tissue interface, motivating the need for an interfacial biosensor. At the same time added material and reagents escalate the risk of impaired biocompatibility and reduced tissue integration. Each surface modification imposes added demands on ensuring safe chemical and mechanical properties.[24]

We characterized these properties with a range of technologies. XPS survey scans of the AuNC-Ti surface confirmed Au-associated peaks indicative of the AuNC coating (**Figure 2a**). The elemental majority of the AuNC-Ti surface consists of Au, which indicates comprehensive coverage of the Ti surface with AuNCs. C1s peaks are also observed presumably from aliphatic contamination. The Ti related peaks in the AuNC-Ti spectra can be attributed to slightly exposed underlying Ti substrate, as penetration depth of XPS is limited to 5-10 nm. O related peaks indicate oxide formation as well. In contrast, survey scans of the bare Ti surface show only Ti related peaks as well as oxide and carbon contamination. The greater presence of oxygen on the bare Ti sample is representative of the exposed $TiO_2$ layer on the bare Ti surface, compared to the gold-coated

AuNC-Ti surface. XPS findings show no unexpected contamination during the AuNC-Ti fabrication process and further verified deposition of AuNCs.

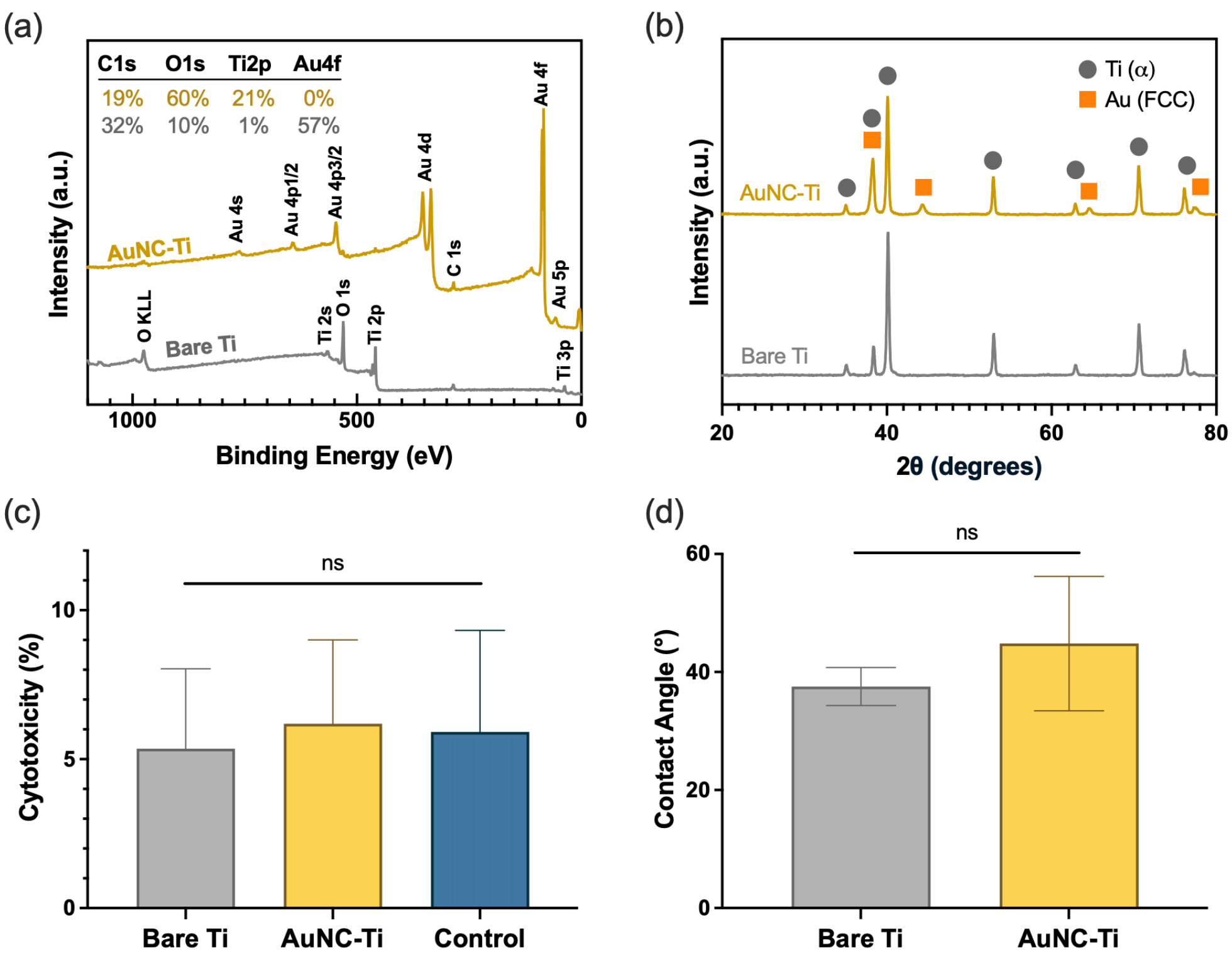

**Figure 2. a)** XPS survey scans over AuNC-Ti surface and bare Ti surface and atomic concentration percentages showing successful coverage of AuNCs on Ti surface. Standard deviations of atomic concentrations can be found in the Supporting Information. **b)** XRD patterns of AuNC-Ti and bare Ti surfaces showing presence of α-Ti along with face-centered cubic Au. The Au (111) diffraction peak overlaps with the Ti (002) peak at ~38 °showing a higher intensity after OAD. **c)** Percent cytotoxicity of bare and AuNC-coated Ti compared to cell and media only negative control showing no statistically significant difference (ns) between all groups. **d)** Contact angle measurements for bare Ti and AuNC-Ti surfaces showing comparable hydrophilicity before and after AuNC coating of Ti. AuNC-Ti demonstrated sufficient wettability for promoting initial protein adsorption and eventual osseointegration.

XRD spectra characterized the crystal structure (**Figure 2b)**. The only dominant peaks on the bare Ti sample are associated with α-phase Ti. After OAD, no change is observed for the α-Ti peaks however there are additional face-centered cubic (FCC) phase Au peaks. The presence of these phases is expected as both these crystal structures are most stable at room and body temperature.[31,32] This indicates that the high temperatures imposed by the OAD process did not alter the phase of the Ti substrate and promoted formation of AuNCs into a stable crystalline phase.

***AuNC-Ti cytotoxicity and wettability assessment***

To confirm that AuNC-Ti surfaces did not impart cytotoxicity to surrounding cells, an LDH release assay was performed after culturing AuNC-Ti and control materials with human aortic endothelial cells (hAECs) for 48 hrs. Percent cytotoxicity, which was assessed through measuring release of LDH accompanying cell membrane lysis, is less than 10% for both bare Ti and AuNC-Ti materials in comparison to the positive control of maximum LDH release (**Figure 2c**). The difference between the amount of LDH released from cells cultured with bare Ti and AuNC-Ti and the amount of spontaneous LDH release in the negative control is not statistically significant ($P > 0.05$). Indeed, Tran *et al.* have already reported that large (> 20 nm), surface immobilized gold nanoparticles (AuNPs) are noncytotoxic and can promote cell survivability in rat cortical embryonic neurons.[33] Others noted that freely dispersed, high aspect ratio AuNPs have no negative impact on affect cell survival, addressing a concern that might arise in the possible event of AuNC detachment.[34] Endocytosis of AuNPs is thought to induce apoptosis, and a study by Arnida *et al.* specifically compared the uptake of Au nanorods and nanospheres in murine macrophages.[35] They observed that Au nanorods were engulfed to a lesser degree than the spherical particles. Their *in vitro* results corroborated the *in vivo* findings where Au nanorods remained in circulation longer than Au nanospheres when injected into mice. It is then possible that the high aspect ratio of AuNCs prevents cellular uptake in hAECs, reducing cytotoxic effects. Longer viability studies can be conducted in future studies to confirm AuNC-Ti biocompatibility over longer time scales (>2 weeks). Considering the importance AuNC morphology in mitigating cytotoxicity, methods to preserve AuNC shape and attachment can be investigated such as protective silica or titania coatings, which are discussed in later sections.

Next, hydrophilicity of the AuNC-Ti material was characterized with static contact angle measurements **(Figure 2d)**. Wettability greatly influences the initial protein adsorption process onto dental and orthopedic implants, which sets forth the complex biological pathways leading to osseointegration.[36] Therefore, characterizing hydrophilicity and surface energy of this material is imperative especially given its dual role as a biomaterial and SERS-active substrate, since Raman enhancement can only occur for biomolecules adsorbed within a few nm to the AuNC surface. No significant statistical difference is observed between the average contact angles of bare nano-smooth Ti and AuNC-coated Ti ($P > 0.05$). The AuNC-Ti surface demonstrates hydrophilicity similar to the bare polished Ti material. Hydrophilic Ti surfaces have previously shown enhanced bone and connective tissue integration resulting of the initial adsorption of key proteins such as fibronectin.[36,37] This adsorption promotes matrix formation, cellular migration, and angiogenesis which are key processes indicative of wound healing we wish to observe with this SERS substrate.[37] Therefore, exhibiting hydrophilicity is promising for both the material's SERS sensing and implant-tissue integration capabilities.

***Spatial mapping of 4-MBA and DTNB on AuNC-Ti surface***

After characterizing biocompatibility, the ability to enhance Raman scattering with spatial resolution was demonstrated using Raman imaging and standard reporter molecules. Spatial mapping of Raman reporter molecules 4-mercaptobenzoic acid (4-MBA) and 5,5'-dithiobis(2-nitrobenzoic acid) (DTNB) was demonstrated by drop-casting 1 μL volumes of each reporter on opposite sides of a sample. Raman confocal images were acquired by scanning over a 13 mm by 13 mm region and collecting 100 measurements (**Figure 3a**). True Component Analysis (TCA), a tool provided by WITec Project Six Software, was used to determine key spectral components using a multivariate analysis-based algorithm to identify the most prevalent spectra.[38] Two distinct spectra (**Figure 3a**) can be identified on the AuNC-Ti surface which were representative of 4-MBA and DTNB. These signals are unlike the low signal to noise ratio (SNR) signals collected from a bare Ti surface coated with the reporter molecules (**Figure S1**). Visualization was enabled by mapping color intensities proportional to the degree in which a component contributed to each measurement. The resulting Raman map was overlayed on an optical sample image to reflect the initial placements of the 1 μL droplets (**Figure 3a**). This demonstrated that AuNC-Ti substrates can enhance Raman signals and provide spatial resolution.

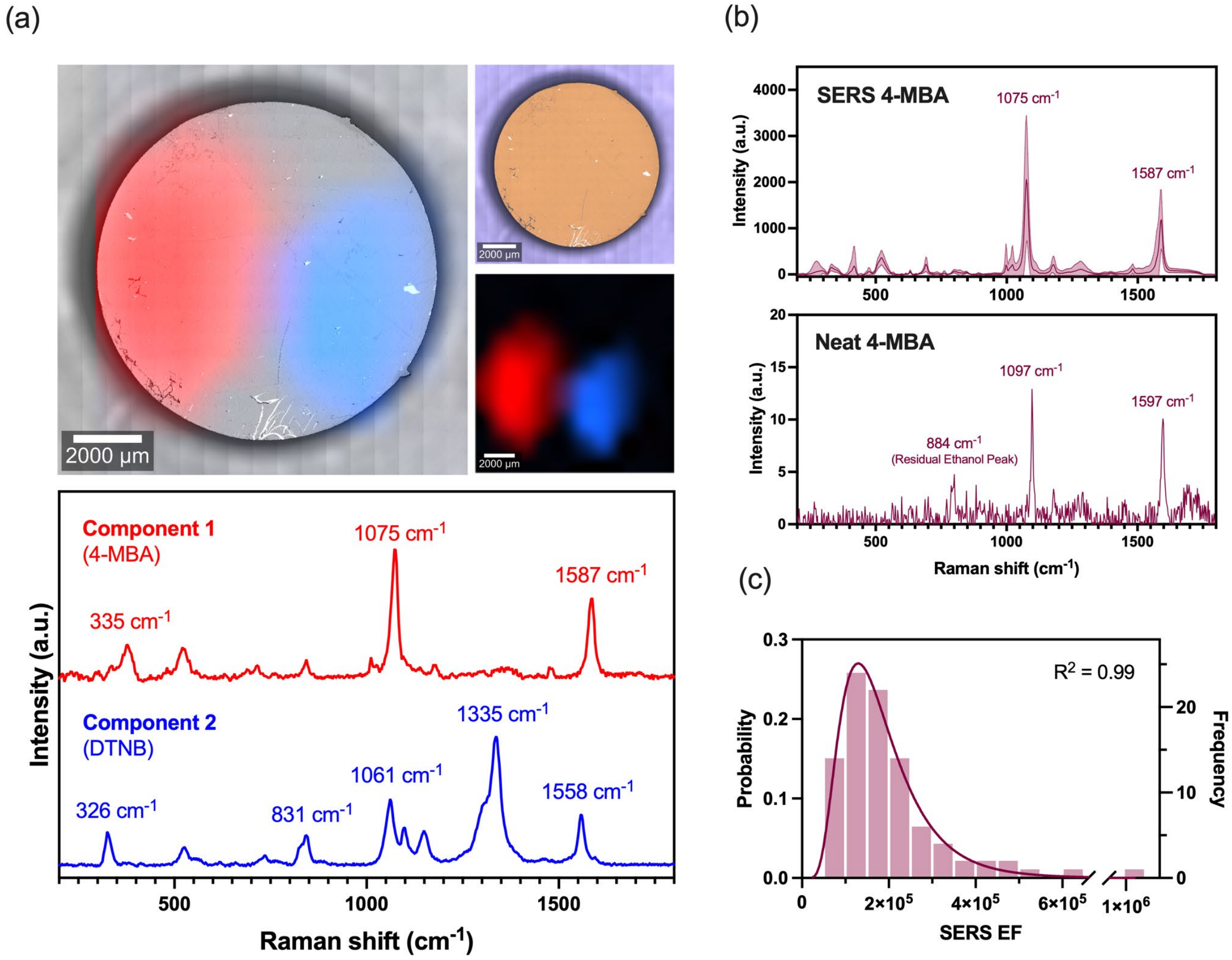

**Figure 3. a)** Spatial Raman mapping of 1 mM 4-MBA (red) and DTNB (blue) drop casted on AuNC-Ti sample with color intensity normalized to the max Raman reporter abundance. Isolated confocal Raman image and underlying sample image shown. Key component spectra determined through True Component Analysis (WITec Project Six) found to represent 4-MBA and DTNB are displayed at the bottom. **b)** SERS spectra of 4-MBA coated on AuNC-Ti surface averaged over n = 93 measurements at varying locations and Raman spectra of neat 100 mM 4-MBA solution in ethanol. **c)** Histogram illustrating distribution of SERS EFS over n = 93 locations on 4-MBA coated AuNC-Ti surfaces and corresponding probability density function. Median EF was calculated to be $1.8 \times 10^5$. The probability density function followed a lognormal distribution through the Anderson-Darling lognormality test (P > 0.05).

### *SERS EF measurements*

The AuNC-Ti surface's ability to enhance Raman signals was quantified via calculating enhancement factor (EF). SERS EF is a theoretically intrinsic value used to measure and compare SERS efficiency but has also struggled with multiple definitions depending on application and experimental conditions. For clarity, the conventional method for calculating SERS substrate EFs (SSEFs) was used with the following equation:

$$SSEF = \frac{I_{SERS}/N_{Surf}}{I_{RS}/N_{Vol}}$$

where $I_{SERS}$ and $I_{RS}$ represent the Raman intensity of the enhanced and unenhanced analyte respectively. [39] $N_{Surf}$ represents the number of adsorbed analyte molecules on the SERS substrate during spectral acquisition, and $N_{Vol}$ represents the number of analyte molecules within the scattering volume without enhancement in the neat solution. More details on the calculations for SSEF with respect to experimental measurements can be found in the Supporting Information.

SERS signature was measured by coating a monolayer of 4-MBA on the AuNC-Ti surface. 4-MBA is a small thiol molecule that forms a well-characterized self-assembled monolayer through Au-thiol bonding, making it an ideal standardized Raman probe molecule for SERS EF quantification.[40] SERS spectra of 4-MBA coated on the AuNC-Ti surface were collected at n = 93 total locations over 5 replicate samples and averaged (**Figure 3b**) and compared with spontaneous Raman spectra obtained from neat 100 mM ethanolic 4-MBA within a glass vial. An EF for each measurement was calculated (**Figure 3c**) with a median of $1.8 \times 10^5$ and interquartile range of $1.3 \times 10^5$. The average SERS enhancement of the AuNC-Ti material reached 5 orders of magnitude consistent with typical SERS substrates.[39]

The SERS EF results followed a lognormal distribution, with a select few EFs approaching 6 orders of magnitude. These outliers likely arise from the variability of AuNC distribution on the Ti surface during deposition which could have been caused by the fusing of neighboring rods, creating smaller than average hotspots. As hotspot size and SERS enhancement are generally inversely related with non-linear effects toward subnano-sized gaps, heterogeneity of gap distance between AuNCs would explain such outliers.[7] Future optimization of the OAD process might mitigate these outliers by altering deposition angle, rate, and thickness to maximize SERS enhancement and enhancement uniformity across the substrate surface. For example, increasing deposition rate and reducing substrate temperature reduces surface diffusion of gold and could promote separate columnar growth.[41] Applying a seed layer composed of regularly ordered surface nanostructures on Ti before OAD has also been demonstrated to control distances between individual nanocolumns.[30] Calibration methods and integrating over area scans can also improve spatial SERS enhancement uniformity through normalization methods described by Nam et al.[42] Further descriptions and demonstrations of this normalization approach can be found in the next section. [42]

***ERS Calibration***

Calibration methods can improve spatial SERS enhancement uniformity through normalization methods described by Nam et al.[42] The relative standard deviation (RSD) of SERS substrate enhancements is a metric used to quantify spatial uniformity of SERS substrate performance. Confocal Raman microscopy allows for measurements to be taken at a specific focal distance, filtering out out-of-plane Raman scattering which allows us to make local measurements at the material-tissue interface. However, ensuring that the SERS substrate topology is always in focus is difficult due to this narrow depth of field, compromising uniformity in spatial SERS signals. Uneven distribution in Raman hotspots can also cause SERS enhancement heterogeneity. Both issues contributed to an average RSD of 15.6% per 100 μm by 100 μm area of the AuNC-Ti material when measuring 4-MBA monolayer signals. The RSD of average Raman intensities of scans over multiple samples was also 36.1%, indicating batch variations. This was addressed through plasmon-enhanced electronic Raman scattering (ERS) calibration methods developed by Nam et al.[42] ERS pseudopeaks in the lower wavenumber domain are characteristic of plasmon resonant metals such as gold and have been demonstrated to act as an internal standard to account for spatial heterogeneity in SERS substrate enhancement.[42] Using this calibration with ERS pseudopeaks around 62 $cm^{-1}$, the relative standard deviation (RSD) of SERS signals within a 100 μm by 100 μm area on substrate was successfully reduced from an average of 15.6% to 7.6% ($P < 0.05$) (**Figure S7**). The RSD of average Raman intensities between scans was also reduced from 36.1% to 6.6% (**Figure S6**). ERS calibration was found to be sufficient in providing excellent uniformity in our material's spatial sensing capabilities, enabling reproducible SERS measurements even in the event of AuNC fusion or other heterogeneities.

This method performed well under benchtop conditions, but future practical limitations must also be considered during translational efforts. ERS calibration alone is yet known to be sufficient for signal normalization in *in vivo* conditions, where larger variations in hot spot distribution can occur due to coating degradation. When paired with more robust coating fabrication methods however, the internal ERS standard could be powerful enough to provide reproducible SERS signals specific to the implant-tissue interface *in situ*.

***Biological tissue characterization at AuNC-Ti interface using multivariate analysis***

We finalized our proof-of-concept study by testing the capability of AuNC-Ti surfaces to characterize tissue content (fresh pork shoulder sourced from a local grocery store) to simulate contact with biological organs. A 40 µm slice of pork shoulder was placed on the sample surface and 1600-pixel Raman maps obtained over a 300 µm by 300 µm area, followed by hematoxylin and eosin (H&E) staining for validation of tissue components. The AuNC-Ti material enhanced signals of probe molecules with large Raman cross-sections, and as well key Raman peaks specific to proteins and lipids at the material interface. Adipose, muscle, and connective tissue could then be identified in Raman maps using machine learning techniques typically used in Raman hyperspectral imaging.[43]

*Principal Component Analysis (PCA)*

PCA is a common dimension reduction methods for multivariate data analysis, and with Raman, for phenotyping and diagnosing pathophysiology on the cellular and tissue level, especially when combined with clustering and classification algorithms.[13,44,45] PCA involves the linear transformation of high dimensional datasets to a simpler coordinate system composed of variables that capture maximal variance, also known as principal components.[43] Its ability to reduce dimensionality and identify key sources of variability has proven to be useful across disciplines including spectroscopic analysis and especially in systems biology. In this study, PCA was used for hyperspectral Raman imaging enabled by decomposing individual spectra into a combination of PCs and projecting them back as a univariate heat map (**Figure 4 (a and b)**).

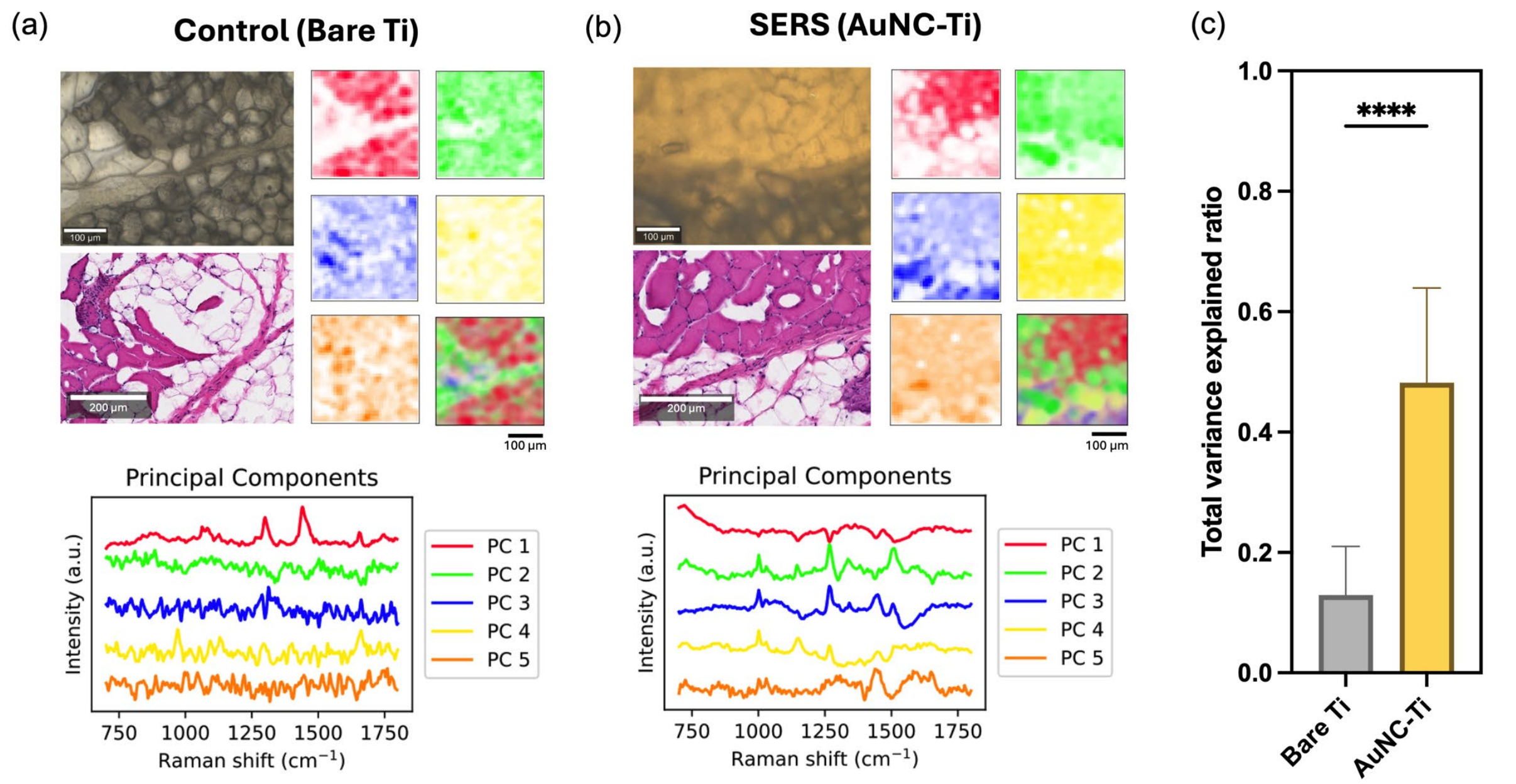

**Figure 4.** Principal component analysis (PCA) of pork shoulder slices laid on bare Ti and AuNC-Ti surfaces. **a)** Control and **b)** SERS Sample images and accompanying H&E-stained images along with projection maps of the first five principal components and merged projection maps with color intensity normalized to maximum component contribution. Corresponding Raman spectra of first five principal components colored consistently with projection maps are shown below. Ability of AuNC-Ti to better enable spatial identification of tissue components (i.e. muscle, adipose, and connective tissue) with Raman imaging is observed. **c)** Comparison between control and SERS group of total variances explained within key components determined using the elbow method quantifying the AuNC-Ti material's ability to amplify spectral information detected during Raman imaging. ($n_{min}$ = 9 Raman scans per group.)

Histology and fresh tissue images were used to identify and locate key tissue components including of muscle, connective, and adipose tissue in each sample. In fresh tissue, muscle fiber cross sections are large and lighter in color while connective tissue is comprised of slightly darker fibrous components. In histology images, muscle fibers appear as dark pink cross sections and connective tissue as pink fibrous areas made up of collagen, fibroblasts, and other extracellular components. The control sample Raman maps distinguish adipocytes within PC1's projection map, but other features such as connective and muscle tissue are less apparent (**Figure 4a**). PC1 includes peaks at 1299, 1441, and 1657 $cm^{-1}$ which are indicative of lipids and adipose tissue.[46] The projection map of PC1 corresponds to the location of adipocytes shown as darker smaller cells in fresh tissue and characteristic large white cells in histology. Key biological peaks are more difficult to identify in PC2 – PC5 due to low signal-to-noise ratio (SNR). Projections of PC2-5 also do not clearly identify the locations of connective tissue and muscle fiber tissue. In contrast to the control, AuNC-Ti sample maps decompose into PCs with higher SNRs and peaks indicative of protein in addition to lipids, including peaks at 1001 (Phe), 1267 (Amide III), and 1508 $cm^{-1}$ (Phe and Tyr) (**Figure 4b**).[47,48] The PCs from AuNC-Ti sample maps loosely denote areas of muscle fibers and connective tissue (PC1 and PC2) and adipose tissue (PC3).

A quantitative comparison on the effectiveness of PCA to capture spectral variances was made between the control and AuNC-Ti maps using the scree or "elbow" method.[49] When performing a PCA, one must select the minimum number of eigenvalues or PCs to extract meaningful variance from a data set and exclude components dominated by noise.[49] The explained variance ratio of the minimum number of PCs can be summed together as the total explained variance ratio and compared between the control and AuNC-Ti groups (**Figure 4c**). Additional details on the elbow method used can be found in the Supporting Information. The AuNC-Ti scans capture 48 ± 16 % of the total variance in comparison to the control scans which only capture 13 ± 8 % of the total variance. Thus, multivariate analysis methods such as PCA can capture more than triple the amount of meaningful variance in a Raman scan of a AuNC-Ti material than the bare Ti control, suggesting that the AuNC coating in conjunction with PCA can enable SERS enhancement of biological signals and improved tissue component detection.

*Vertex Component Analysis (VCA)*

Additional qualitative analysis was performed to produce more biologically interpretable components and abundance maps (**Figure 5**), as PCA provides a quantitative comparison but does not necessarily produce physically relevant component spectra due to mathematical constraints of the method (e.g. the presence of negative intensity peaks in PC1 and PC3 for AuNC-Ti tissue scans (**Figure 4b**)).[43] We then used a more robust unmixing method commonly employed in hyperspectral image analysis known as VCA, which extracts positive endmembers by estimating the most "pure" pixels in a hyperspectral dataset analogous to PCs.[50] A key advantage of VCA however is its ability to produce endmembers composed of only physically possible spectra unlike components from PCA.[50] VCA allows for improved characterization through peak identification of key endmembers compared to PCs in relation to tissue components.

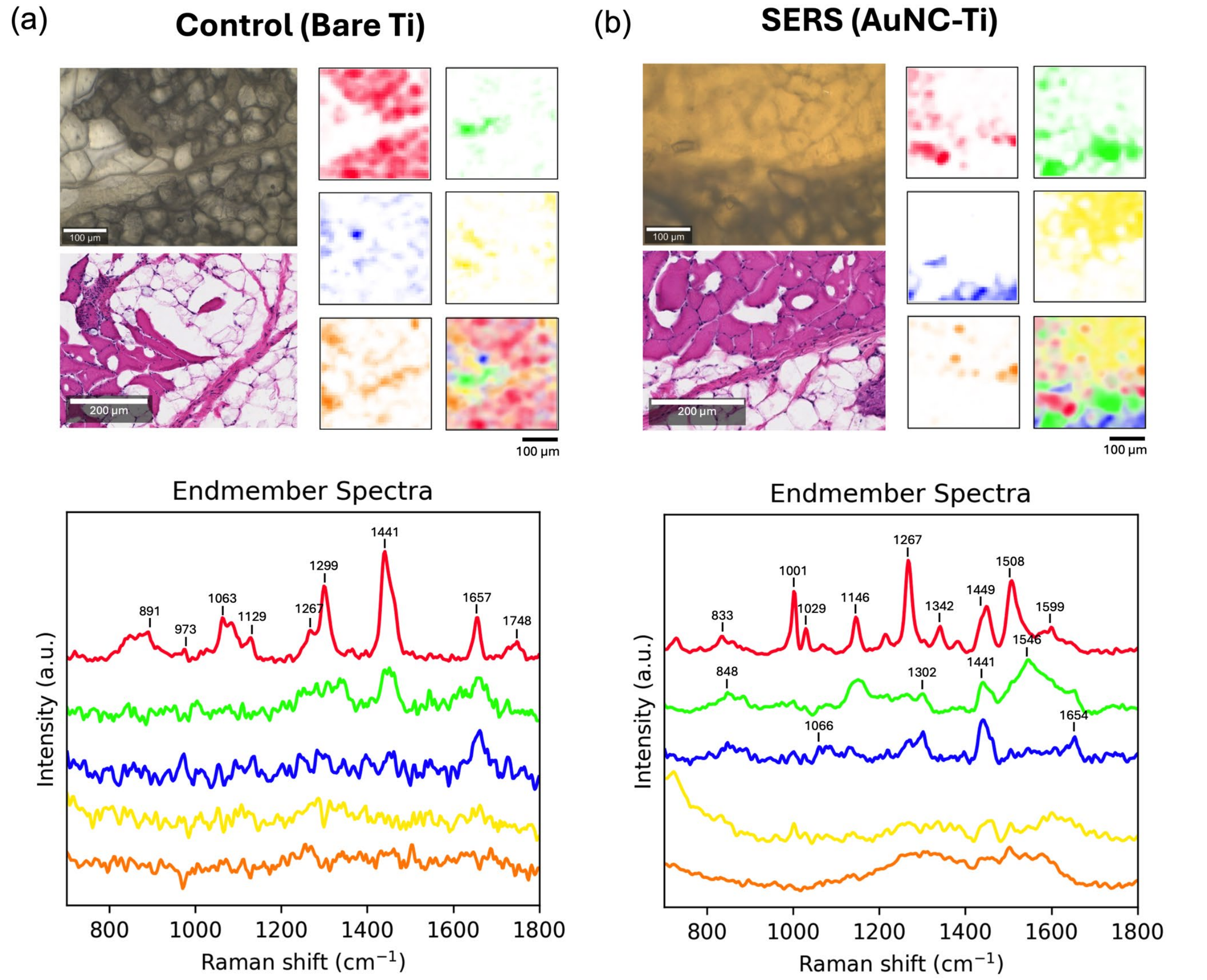

**Figure 5.** Vertex component analysis (VCA) performed on pork shoulder slices laid on bare Ti and AuNC-Ti surfaces. **a)** Control and **b)** SERS Sample images and accompanying H&E-stained images along with endmember abundance maps and merged maps of endmembers, Color intensity is normalized to maximum endmember abundance. Corresponding Raman spectra of first five principal components colored consistently with abundance maps are shown below with key peaks labeled, indicating that the AuNC-Ti material promotes detection of a greater number of key peaks representative of biological components compared to the control.

In cases where substances are not present in their purest form, i.e. they are comprised of multiple components which is typical of biological samples, it must be noted that endmember estimates do not necessarily represent pure substances. Interestingly, VCA can identify key Raman spectra indicative of the pork tissue elements. Endmember spectra identify known tissue components (connective, muscle, and adipose tissue, amorphous carbon peaks due to unintentional burning of gold, and background) which were verified through histology and optical imaging. The burning is likely a consequence of especially thin tissue or exposed substrate sections allowing laser induced damage to occur. Photothermal burning is expected to be less of an issue *in vivo* as tissue thickness would not vary as extremely, allowing laser power settings to be more easily tuned for optimal tissue-material interface excitation while mitigating photothermal heating. In this case, only five endmembers are needed, and additional endmembers were generally observed to be noise or redundant.

Unlike the PCA results, the endmembers identified through VCA are specific to either adipose, muscle, or connective tissue and occasional mixtures. In the control group, only adipose is clearly identifiable in Endmember 1 and proteins to a slight degree in Endmembers 2-5 (**Figure 5a**). This is unsurprising as lipids are rich in highly polarizable C-H and C-C bonds, unlike other biochemicals such as proteins.[51] Adipose tissue is clearly identified in abundance maps of Endmember 1, and muscle and connective tissue to a degree in Endmember 2 and 5 maps respectively. Endmembers 2-5 struggle with low SNRs however, making component identification and abundance mapping of muscle and connective tissue unclear.

The AuNC-Ti maps in contrast have multiple endmembers indicative of both proteins and lipids and to a degree nucleic acids and polysaccharides (**Figure 5b**). Endmember 1 contains strong peaks characteristic of proteins such as collagen, suggesting the presence of connective tissue at the interface between muscle and fat tissue (assignments in **Table 1**). Peaks at 728 $cm^{-1}$ and 848 $cm^{-1}$ are also attributed to carbohydrates and nucleic acids respectively unlike in the control.

Endmember 2 contains peaks characteristic of both lipids and proteins, implying that a mixture of both was detected at the interface. Endmembers 3 and 4 are indicative of lipids and proteins respectively, and their abundance maps are analogous to the location of adipose and muscle tissue**.** Endmember 5 is likely a detection of amorphous carbon D and G bands due to burning gold. All abundance maps from the AuNC-Ti material match the area relative to tissue components found via histology.

**Table 1.** Spectral peak assignments of pork shoulder tissue found via VCA.

| Peak Position ($cm^{-1}$) | | | |
|---|---|---|---|
| **Control** | **AuNC-Ti** | **Tentative Peak Assignment** | **Reference** |
| | 728 | Adenine | [52,53] |
| | 833 | Phe, Tyr | [47,52] |
| | 848 | Tyr, Pro, polysaccharides | [54] |
| 891 | | Gly | [45] |
| | 1001 | Phe | [52] |
| | 1029 | Phe, C-C skeletal stretch | [52,53] |
| 1063 | 1068 | C-C skeletal stretch, palmitic acid, Pro | [52,54] |
| 1129 | | C-C skeletal, C-N stretch | [52,53] |
| | 1146 | C-C stretch, COH deformation | [52,53] |
| | 1213 | Phe, Tyr | [52] |
| 1267 | 1267 | Amide III, CH2 deformation (lipids), Phe, Tyr | [47,52,54] |
| 1299 | 1302 | Palmitic acid (lipids) | [54,55] |
| | 1342 | CH bend (in residues) | [52] |
| 1441 | 1441 | CH2 scissor (lipids) | [46,54] |
| | 1449 | CH3, CH2, CH bend (proteins) | [52,53] |
| | 1508 | Phe, Tyr (collagen) | [47] |
| | 1546 | Amide II | [52] |
| | 1599 | Phe | [52] |
| 1657 | 1654 | C=C stretch, Amide I | [46,52] |
| 1748 | | C=O stretch | [46,54] |

The improved SNR of Endmember 1 in the SERS group (**Figure 5b)** may be due to the connective tissue being closer in proximity to the AuNCs in comparison to other components, as other SERS samples display stronger lipid signals or vice versa. Additional tissue maps are available in the Supporting Information (**Figure S17 and S18)**. The control group consistently exhibits only lipid specific signals in all samples due to the lack of enhancement and dominance of surface scattering, demonstrating the selectivity of the AuNC-Ti surface in enhancing only molecules in proximity of hotspots at the interface.

The AuNC-Ti material has much increased the capacity to reflect on tissue elements than control materials, displaying more Raman peaks especially those classically representative of peptide and lipid characteristics (**Table 1**). The intensity, width, shifts, and combinations of these peaks can then provide insight into biochemical composition and structure of the embedding milieu and when used *in vivo*.[45,52] The peaks detected are sensitive to peptide conformational changes, residue composition, pH, and many other factors that grant highly multiplexed sensing capabilities.[53,54,56,57] Previous research has even demonstrated the potential of SERS to identify misfolded proteins and reconstruct entire peptides under ideal conditions.[58,59]

Measuring the complex biological milieu however is carried out under conditions far from the ideal for straightforward Raman spectra interpretation. Machine learning and artificial intelligence have allowed us to overcome this challenge with multivariate analysis and deep learning's affinity to analyze high dimensional datasets. A study by Sugiyama et al. demonstrated this through their ability to parse ECM components with multivariate analysis down to collagen, nuclei, lipids, elastic fibers, aggrecan, versican, and residual proteins in Raman images of aorta cross sections.[60] Our findings corroborate these results and confirm the effectiveness of SERS to enhance a variety of interfacial Raman signals, improving the performance of multivariate analysis in tissue component unmixing. The AuNC-Ti material enables us to detect 17 peaks relevant to lipids, proteins, and nucleic acids in pork shoulder tissue, compared to only 8 in the control group which are largely only indicative of lipids (**Table 1**). We can detect key peaks at 1001, 1029, 1213, 1508, and 1599 $cm^{-1}$ to distinguish collagen rich connective tissue, which envelops implants when osseointegration fails to occur. Spectra of connective tissue are relatively stronger than spectra of other proteins in pork shoulder tissue, likely because of the high density of collagen fibrils. This allows phenylalanine (Phe) and tyrosine (Tyr) peaks (particularly at 1001 and 1029 $cm^{-1}$) to dominate in dense connective tissue spectra as a result of their aromatic functional groups.

In the context of tracking implant-tissue interactions, detection of dense proteins could be instrumental in defining fibrotic tissue formation in the face of micromotions and inadequate integration. If the initial stage of fibrosis is caught early, a clinician could recommend methods to control micromotions which are visually unperceivable, such as eating softer foods during the healing process of a dental implant. Healing in turn can be tracked by identifying the initial fibrin matrix formation, migration, differentiation, and proliferation of osteogenic and immune cells, and finally de novo bone formation and calcification. A previous study by Jain et al. identified stages of healing in mice (inflammation, proliferation, and hemostasis) through multivariate factor analysis, non-negative least squares fitting, and clustering, and Garg et al. similarly developed a SERS-active textile-based bandage to track bacterial infections with machine learning.[61,62] We hypothesize that similar tracking of complications and healing progress can be achieved with SESORS for implant healing since we can obtain multiplexed spectral data at the tissue-material interface with nondestructive means using a SERS-active implant material. With the use of machine learning and SESORS, spectral data from the implant surface *in vivo* could be classified into various pathophysiological states, allowing for immediate monitoring capabilities for clinicians and improved peri-implant healing.

## NEXT STEPS AND FUTURE OUTLOOK

This ex vivo study provided pivotal insights into the capability of SERS substrates in sensing biochemical characteristics at the tissue-implant interface, but multiple considerations in material design, signal acquisition, and data analysis should also be made when translating to an *in vivo* model. Unlike in the ex-vivo study, the coating must withstand the *in vivo* environment for at least multiple days, ideally months, to best monitor the healing process of an implant.[63] This includes exposure to pH ranges between 5.5-7.4, sustained temperatures around 37°C, gold-reactive halides, oxidative species, mechanical stresses, fluid movement, and tissue remodeling processes.[17,64] These processes are expected to detach or degrade AuNCs overtime causing AuNCs to shorten or fuse with one another, and therefore, methods to protect the AuNC coating from the harsh environment without sacrificing sensing capabilities should be explored. The *in vivo* environment will also require deep tissue sensing abilities, and most importantly, involves dynamic spatial and temporal processes that should be sensed by the SERS platform.

There are multiple methods to address coating stability and mechanical robustness for preserving AuNC surface morphology *in vivo*, which strongly influence both SERS enhancibility and biocompatibility as suggested by previous literature.[7,35,65]. Ultrathin oxide coatings on Ag nanorod (NR) arrays and colloidal AuNPs have been demonstrated to provide stability in aqueous solutions by preventing oxidizing agents and reactive ions from causing delamination or degradation of nanostructures while preserving SERS activity.[66,67] A silica or titania coating can be similarly applied to AuNC-Ti surfaces, preventing shortening, fusing, or detachment of AuNCs which can compromise SERS enhancibility and biocompatibility. Ti adhesion layers, which are 2-5 nm thick layers of Ti, are used to physically bind metal and nonmetal surfaces and can also prevent AuNC detachment. Despite the substrate already being composed of Ti, a $TiO_2$ surface layer forms when exposed to oxygen in the air, weakening adhesion of AuNCs to Ti implant surfaces. Previous literature demonstrates Ti adhesions layers to be beneficial during OAD processes to attach SERS-active nanostructures to oxide-coated substrates[68], and this method is promising for biomedical applications considering the inherent biocompatibility of Ti. A thin Ti adhesion layer can be deposited on the oxidized Ti substrate before AuNC deposition under high vacuum, preventing surface oxidation and allowing for improved subsequent adhesion of AuNCs. Another potential method to improve coating stability could involve covalent bonding of AuNPs to $TiO_2$ surfaces. Despite covalent bonding of AuNPs requiring more complex synthesis methods, the strength of AuNP attachment to $TiO_2$-Ti surfaces could potentially outperform attachment of OAD fabricated coatings. A method has already been demonstrated *in vivo* on Ti implants coated with AuNPs to promote osseointegration by Heo et al., generating confidence in biocompatibility and interest in application to SERS.[19]

Biocompatibility in regard to bone tissue specifically can also be further addressed using long-term viability and osteogenic induction studies using mesenchymal stem cells and osteogenic cell lines.[69] Future *in vitro* and *in vivo* studies using a spatially offset detector system can be performed to demonstrate surface sensing selectivity beneath tissue, which has been demonstrated to be effective in sensing analytes up to 8 mm within bone.[70,71] Laser power and exposure time could also be tuned to maximize SERS sensitivity while mitigating potential photothermal heating related damage. This can be done by estimating implant depth by correlating spatial offset distance with Raman intensity of an internal standard such as the ERS pseudo peak, which can inform optimal laser power settings.[72] The effect of laser incidence angle can also be studied, as *in vivo* sensing seldom allows perfectly normal signal acquisition that was possible *in vitro.* Possible fluctuations in SERS enhancement may occur at varying incidence angles and orientations, and solutions such as ERS calibration to normalize differences in plasmonic resonance could be investigated.

Capturing the dynamic processes at the implant interface may also require hotspot regeneration techniques in the event of unwanted protein fouling. While protein composition on the implant surface is dynamic and a major factor in implant integration, the exact specificity of protein adsorption at an AuNP-modified implant surface is still be investigated, Ideally, the SERS-active implant should be able to sense the turnover of proteins starting from serum proteins and molecules followed by cell attachment and osteogenic differentiation, matrix formation, and de novo bone formation. Unwanted protein adsorption leading to biofouling and implant failure should also be sensed, indicating the need for clinical intervention. In the event of unspecific surface adsorption limiting sensing capabilities over time, active anti-fouling methods can be explored such as electrochemical or photocatalytic hotspot regeneration which temporarily removes analytes from the substrate surface.[73,74] These methods allow for regeneration to be controlled and spatially specific since protein surface adsorption is also crucial to implant success, and disturbance to necessary implant-tissue interfacial proteins should be minimized. Conversely, unwanted protein fouling could be addressed by active surface protein degradation via electrochemical or photocatalytic methods, potentially being a therapeutic intervention to aid implant integration.

The complexity of developing a self-monitoring orthopedic or dental implant leaves many opportunities for additional engineering and investigation, but this initial SERS-based proof-of-concept motivates and informs further maturation of this technology. There is exciting potential to make significant advances in implant technology and diagnostics, and it will likely require highly interdisciplinary efforts involving materials scientists, engineers, optical physicists, biologists, clinicians, and machine learning experts.

## CONCLUSION

We developed a SERS-active implant using AuNC coating on Ti, a common implant material. Material characterization confirmed minimal cytotoxicity and adequate Raman enhancement factors up to five orders of magnitude with excellent spatial uniformity (< 10% RSD) in SERS enhancement using ERS pseudo peak normalization. We quantitatively and qualitatively demonstrated that the material can enhance tissue specific signals indicative of biochemicals to characterize connective, adipose, and muscle tissue with spatial resolution, specifically at the surface-tissue interface using PCA and VCA. The AuNC-Ti material increase explained variance in key PCs from 13 ± 8 % to 48 ± 16 % and more than doubled the number of biologically relevant peaks found using VCA compared to the control. SERS-active coatings on implant materials hold great promise as an *in vitro* platform to study the temporal and spatial dynamics of tissue-material biology, as well as the foundations for an *in vivo* self-sensing material for implant tissue state tracking. Since multivariate analysis was able to distinguish muscle, connective, and adipose tissue in contact with AuNC-Ti surfaces, this motivates future applications in bone tissue. Such translations can be achieved by optimizing AuNC-Ti material for coating stability and biocompatibility by exploring Au deposition parameters, protective ultrathin oxide coatings, and other fabrication methods outside of OAD to enable long-term studies *in vivo*.[27,30,75]

The idea of "smart" implants has long been explored - we now view it through the lens of computational biology, incorporating the detection and interpretation of highly multiplexed real-time signals. We posit that nanotechnology such as SERS-active coatings can provide insight into a multitude of simultaneous biological metrics, compared to the limited nature of current smart implants, to improve our understanding of implant biology. Just as systems biology has spanned across an increasing number of –omes, great promise lies in collecting and studying spectral data, advancing knowledge in diagnostics in a nondestructive and multiplexed manner.

We consider this study the first step toward applying "spectromics" to medical implant biology and engineering.[76] When combined with SESORS, we anticipate that implant materials can be engineered to become self-sensing to allow monitoring of pathophysiological state in neighboring tissue, creating new avenues in the field of personalized medicine.

## METHODS

### *Fabrication of AuNCs on Ti substrate*

AuNCs were deposited on top of Ti surfaces using the oblique angle deposition (OAD) method. Grade 2 Ti discs of 10 mm diameter and 1.02 mm thickness were used as the substrate. Nano-smoothness was first achieved by grinding the Ti surface using 300 grit silica-carbide grinding paper after mounting to an automated grinder (Struers Tegrapol) under 6 N load at 300 rpm with water cooling until planar. Ti disks were then polished (Buehler AutoMet 250) at 150 rpm and 6 N load using a 9 μm diamond suspension and UltraPad polishing cloth (Buehler) for minimum of 10 min. Finally, the surfaces were polished with 0.02-0.06 μm colloidal silica on a ChemoMet polishing cloth (Buehler) at 150 rpm and 5 N load for 10 min or until achieving an optical mirror finish. Reproducibility of nano-smoothness was achieved consistently between batches. Please see **Figure S2** for a comparison between polished and unpolished Ti surfaces under SEM. Ti surfaces were cleaned prior to AuNC deposition through a sequence of ultrasonic baths in acetone, isopropanol, and deionized water and then dried overnight.[77] The cleaned substrates were then placed in an electron-beam evaporator system (AJA International Inc.) at an incident beam angle of 87° to the normal of the Ti substrates. Au was deposited under high vacuum at a base pressure of 1e-5 torr onto Ti substrates using OAD with a deposition rate of 3 Å/s until a nominal thickness of 500 nm was achieved. The deposition rate and thickness of the gold film was monitored using a quartz crystal microbalance.

***Surface characterization***

The surface morphology was characterized using scanning electron microscopy (SEM, Zeiss Gemini 450) and imaged at 2 kV. ImageJ was used to estimate nanocolumn dimensions, surface density, and angle. AuNC angle was estimated by tilting the sample at a 45° angle during SEM imaging and performing tilt compensation. Chemical compositions of the surfaces were measured using X-ray photoelectron spectroscopy (XPS, PHI VersaProbe II). Survey scans were taken from 0 to 1100 eV at 15 kV with a takeoff angle of 45°. Peak analysis was performed using MultiPak and CasaXPS. Crystal structure and composition were characterized using X-ray diffraction analysis (XRD, Rigaku SmartLab). Pattern analysis was conducted using the HighScore software. Contact angles were measured using a Ramé-Hart Model 500 Advanced Goniometer and DROPimage Advanced software. 7 μL of deionized (DI) water droplets were placed upon each sample surface after rinsing with ultra-pure (UP) water and drying under a nitrogen stream.

***SERS performance measurements***

*Calculating enhancement factor*

Confocal Raman microscopy (Oxford Instruments, WITec alpha300 apyron) was used to characterize the surface's ability to enhance Raman signals of analytes in contact. 4-mercaptobenzoic acid (4-MBA, Sigma-Aldrich), a Raman probe with well characterized Raman spectra, was used to characterize enhanceability of the coating. Substrate samples were incubated in 1 mM 4-MBA ethanolic solution for a minimum of 1 hr before rinsing with ethanol 3 times and left to dry. Raman measurements were taken at a minimum of 14 random locations on each substrate over 5 replicate substrates totaling to n = 92 overall measurements. A 785 nm laser was used for spectra acquisition at a power of 5.74 mW, 0.5 sec integration time, 300 g/mm grating, and 20x magnification. 20 accumulations were taken at each location. Raman measurements were made of neat 4-MBA solution over 5 replicates for EF calculations which are detailed in the Supporting Information. Cosmic ray removal and background subtraction were performed using the WITec Project software.

*Spatial Raman mapping with 4-MBA and DTNB*

1 μL of 1 mM 4-MBA and DTNB (Thermo Fisher Scientific) in ethanol were respectively drop-casted on opposite sides of the sample to create two thin films on the AuNC-Ti surface. This analysis was done on each sample after performing a 13 mm by 13 mm scan over the entire sample area. Each scan consisted of collected spectra taken on a 10 x 10 grid across the scan area. Raman scans were taken using a 785 nm laser at a power of 5.74 mW, 0.5 sec integration time, 300 g/mm grating, and 10x magnification. True Component Analysis (TCA) available in the WITec Project software was used to determine key spectra representative of 4-MBA and DTNB using spectral unmixing after cosmic ray removal and background subtraction. Spectra of the two Raman probes were mapped with a representative color and intensity proportional to the normalized amount of key spectra contribution to each pixel. Color maps were overlayed with corresponding sample images.

*ERS Calibration*

Raman scans were performed over 100 µm by 100 µm regions consisting of 25 total measurements each over 4-MBA coated samples to assess homogeneity of Raman enhancement. Laser parameters were consistent with those described earlier but at 20x magnification. Three scans were performed on each triplicate sample at varying locations, resulting in 225 total measurements per group.

***Cell culture***

Human aortic endothelial cells (HAECs, PromoCell) were cultured in endothelial cell basal medium-2 (EBM-2, Lonza) with supplement kit (CC-4176, Lonza) and 1% (v/v) penicillin/streptomycin. HAECs were incubated at 37°C and 5% CO2 in 75 cm2 cell culture flasks. Cells were expanded until passage 6 after changing growth media every 2-3 days until reaching 90% confluence before trypsinization. Prior to cell seeding, all substrate surfaces were sterilized via incubating in 70% ethanol for 5 min, rinsing with sterile PBS 3 times, and a 30 min incubation in sterile PBS under UV light. The sterile substrate surfaces were seeded with HAECs resuspended to a concentration of $2.0 \times 10^4$ cells/mL.[77]

### *Cytotoxicity evaluation*

The cytotoxicity of each Ti substrate with various surface treatments was characterized by measuring lactate dehydrogenase (LDH) release using the CytoTox 96® Non-Radioactive Cytotoxicity Assay (Promega) according to the manufacturer protocol. After culturing HAECs on each substrate for two days, 50 μL aliquots of growth media supernatant from each sample was collected and placed into a 96-well plate. Maximum LDH release controls were prepared by adding 10% (v/v) Triton X-100 to sample wells culturing cells without substrates. Aliquots (50 μL) of lysate were placed into the 96-well plate after 45 min of incubation. Vehicle controls to measure spontaneous LDH release were similarly prepared using aliquots of supernatant from sample wells containing only cells without lysing. Aliquots of only cell culture medium was finally prepared to later determine the culture medium absorbance background. 50 μL of assay reagent was added to each well containing test and control sample aliquots and incubated at room temperature for 30 min while protected from light. 50 μL of stop solution (1M acetic acid) was added to each well after incubation, and absorbances were measured at 490 nm with a plate reader (Thermo Fisher Scientific). The absorbances of LDH release were calculated after subtraction of cell culture background absorbances. Percent cytotoxicity was calculated as experimental LDH release absorbance over maximum LDH release absorbance.

### *SERS mapping of porcine tissue at AuNC-Ti interface*

Pork shoulder was obtained from a local grocery store, frozen, and sliced across the grain. Samples were obtained using a 10 mm biopsy punch and snap frozen in optimal cutting temperature (OCT) compound. Samples were then cryo-sectioned into 40 μm thick slices for Raman imaging and 10 μm thick slices for hematoxylin and eosin (H&E) staining. The 40 μm slices were placed on a quartz coverslip and washed in PBS to remove the OCT compound. The tissue slice was then placed above an AuNC-Ti sample with the coverslip above before Raman imaging. Large area Raman scans were performed over a 300 µm by 300 µm area consisting of 40 by 40 pixels (Oxford Instruments, WITec alpha300 apyron). Scans were obtained using a 785 nm laser at a power of 30 mW, 0.5 sec integration time, 300 g/mm grating, and 20x magnification

***Multivariate analysis of tissue Raman maps***

RamanSPy, an open-source Python library developed by Georgiev *et al.,* was used for preprocessing and subsequent principal component analysis (PCA) decomposition and vertex component analysis (VCA) unmixing.[78] Preprocessing consisted of cropping to the finger-print region (700-1800 $cm^{-1}$), de-spiking by filtering using a modified z-score[79], denoising using a Gaussian filter, baseline correction via Asymmetric Least Squares, and vector normalization. In addition to cosmic ray removal, de-spiking was also used to remove unidentified contaminant peaks by reducing the threshold parameter. Otherwise, the threshold remained constant for the uncontaminated scans.

PCA was performed by generating the first nine principal component spectra and projection maps for each scan. The first five projection maps were visualized and merged. The explained variance ratio was calculated for each component and plotted over increasing principal components for each scan. The elbow method was used to determine the minimum number of principal components needed to describe the majority of variance in a scan. This was determined by finding the cutoff where the change in explained variance was less than 10% of the maximum explained variance of all components. Components before the cutoff were used to calculate "total variance explained" of each scan by summing the explained variance ratios for each component within the cutoff threshold. Example elbow plots can be found in the Supporting Information.

VCA was performed by generating the first five endmembers of each scan and plotting their spectra and abundance maps. Abundance maps were merged for visualization purposes. Peak identification was done using RamanSPy.

***H&E staining***

10 μm thick tissue samples were placed on charged glass slides, washed in PBS, and fixed with 10% neutral buffered formalin for 10 min. Samples were then washed in tap water before H&E staining using a Sakura stainer/coverslipper. Whole slide images were obtained using a Leica Aperio Slide Scanner at 20x magnification.

***Statistical analysis***

XPS, XRD, and SEM analyses were performed on $n_{min}$ = 2 replicate samples, and Raman imaging was performed on $n_{min}$ = 3 replicates. During SEM imaging, 3 images were taken at random across the substrate surface per magnification level. AuNC dimensions and angles were evaluated at n = 9 locations (3 per replicate sample). AuNC surface density was also estimated over n = 9 separate 1 $\mu m^2$ areas. XPS analysis was performed at n = 2 point locations on each replicate substrate. Contact angle measurements were taken on n = 6 replicate samples for each group after rinsing with DI water and drying with nitrogen.[80] Cytotoxicity studies were carried out on n = 4 replicates for control and test samples. For biological tissue analysis, $n_{min}$ = 3 maps were taken at various locations on each tissue covered sample. Maps were acquired on n = 5 AuNC-Ti coated samples and n = 4 bare Ti surfaces as control. Tissue sister slices taken for H&E staining were done in n = 3 replicates. A Welch's T test was used to compare contact angle measurements before and after AuNC deposition as well as explained variances of key principal components during PCA of porcine tissue SERS maps ($P < 0.05$ = *, $P < 0.01$ = **, $P < 0.001$ = ***). Analysis of variance (ANOVA) and post-hoc Tukey's Test was performed for multiple comparisons of percent cytotoxicity, SERS intensities, and SERS relative standard deviations (RSD) ($P < 0.05$ = *, $P < 0.01$ = **, $P < 0.001$ = ***, and $P < 0.0001$ = ****). Statistical analysis was performed in GraphPad. Uncertainties were reported as standard deviations.

## SUPPORTING INFORMATION

Raman spectra of 4-MBA and DTNB coated bare Ti, characterization of AuNC-Ti material using unpolished Ti substrate, ERS calibration results, finite-difference time-domain (FDTD) results investigating various AuNC-Ti lengths and laser incidence angles, AuNC-Ti reflectance spectra, details regarding SSEF calculations, and additional tissue VCA results (PDF).

## ACKNOWLEDGEMENTS

We would like to thank Dr. Jeon Woong Kang, Prof. Mercedes Balcells, and Dr. Rebecca Zubajlo for their expertise, experimental design suggestions, and supply of HAECs. We would also like to acknowledge the facilities and technical support of MIT.nano, Koch Institute's Robert A. Swanson (1969) Biotechnology Center (specifically the Peterson (1957) Nanotechnology Materials and Hope Babette Tang (1983) Histology Core Facility), the Institute for Soldier Nanotechnologies, and MIT Laboratory for Physical Metallurgy in material fabrication and characterization. This research was supported by the NSF Graduate Research Fellowship.

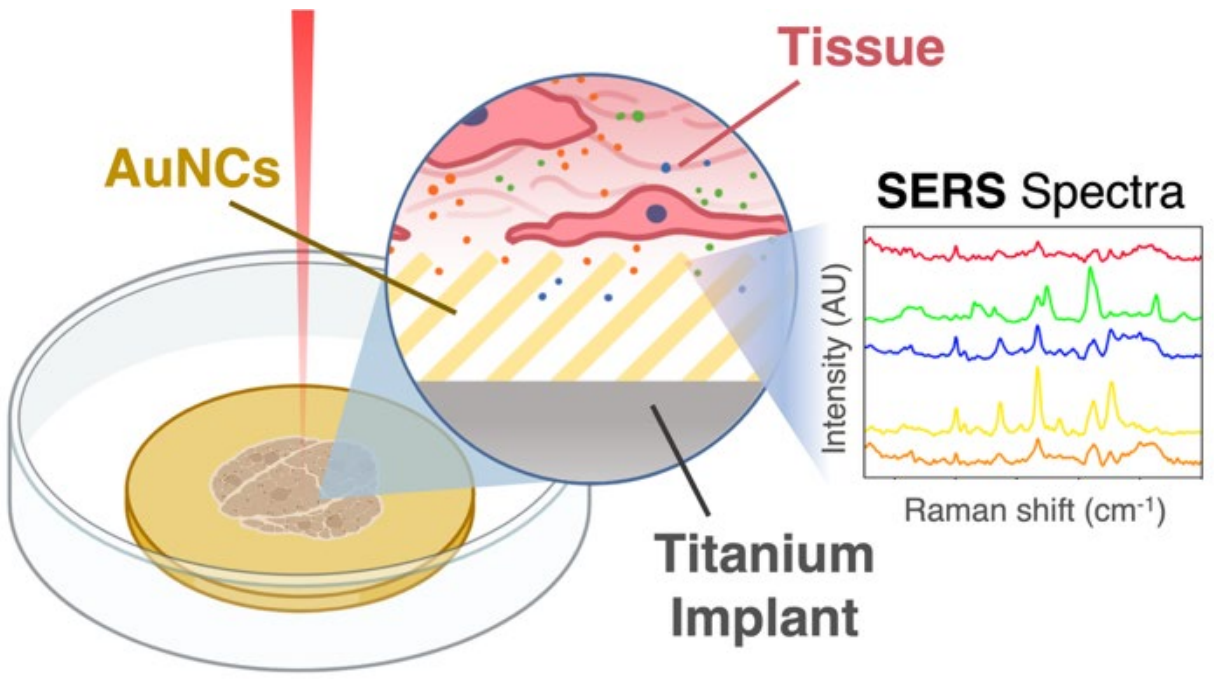

**For Table of Contents Only**

# **Supporting Information:** Nano-engineered surface enhanced Raman spectroscopy substrates for probing tissue-material interactions

*Connie M. Wang[1]*, Roberta M. Sabino[2], Aditya Garg[3], Ahmed E. Salih[3], Loza F. Tadesse[3,6,7]*, Elazer R. Edelman[4,5]*

[1]Department of Biological Engineering, MIT, Cambridge, MA 02139, USA

[2]Department of Chemical and Biomedical Engineering, University of Wyoming, Laramie, WY 82071, USA

[3]Department of Mechanical Engineering, MIT, Cambridge, MA 02139, USA

[4]Institute for Medical Engineering and Science, MIT, Cambridge, MA 02139, USA

[5]Cardiovascular Division, Brigham and Women's Hospital, Harvard Medical School, Boston, MA, 02115, USA.

[6]Ragon Institute of MGH, MIT and Harvard, Cambridge, MA, USA

[7]Jameel Clinic for AI & Healthcare, MIT, Cambridge, MA, USA

* **Corresponding author (email):** Connie M. Wang (cwang1999@gmail.com), Loza F. Tadesse (lozat@mit.edu)

## SUPPLEMENTARY FIGURES

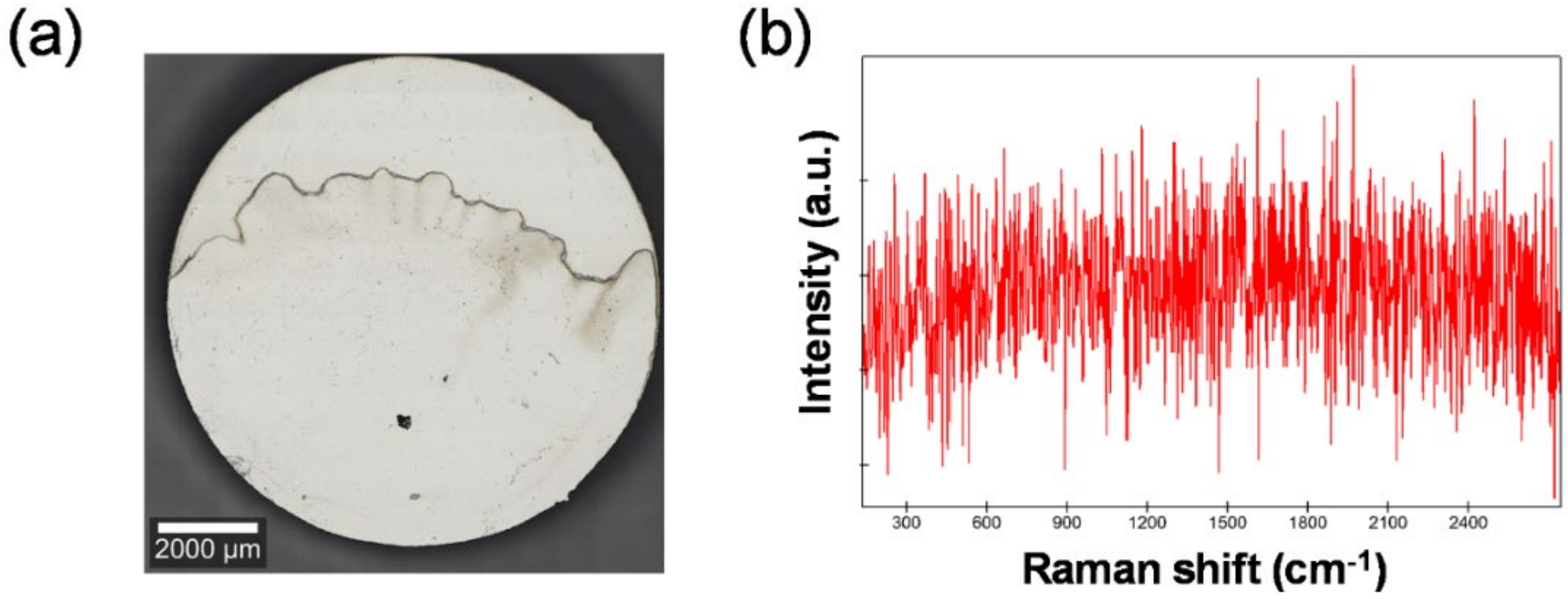

**Figure S1. a)** Polished control bare Ti substrate with 1 mM 4-MBA and DTNB drop casted at opposite end and **b)** representative surface Raman spectra. Both reporter coatings exhibited spectra similar to background noise.

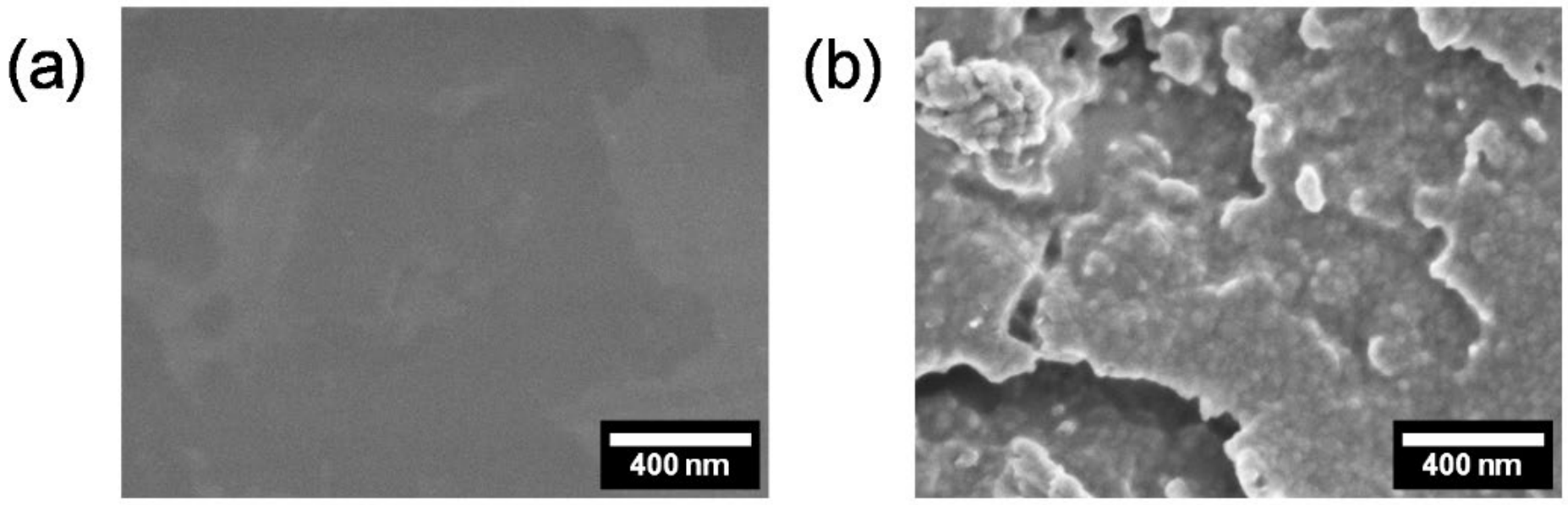

**Figure S2.** SEM image of bare Ti substrate **a)** polished and **b)** unpolished (rough).

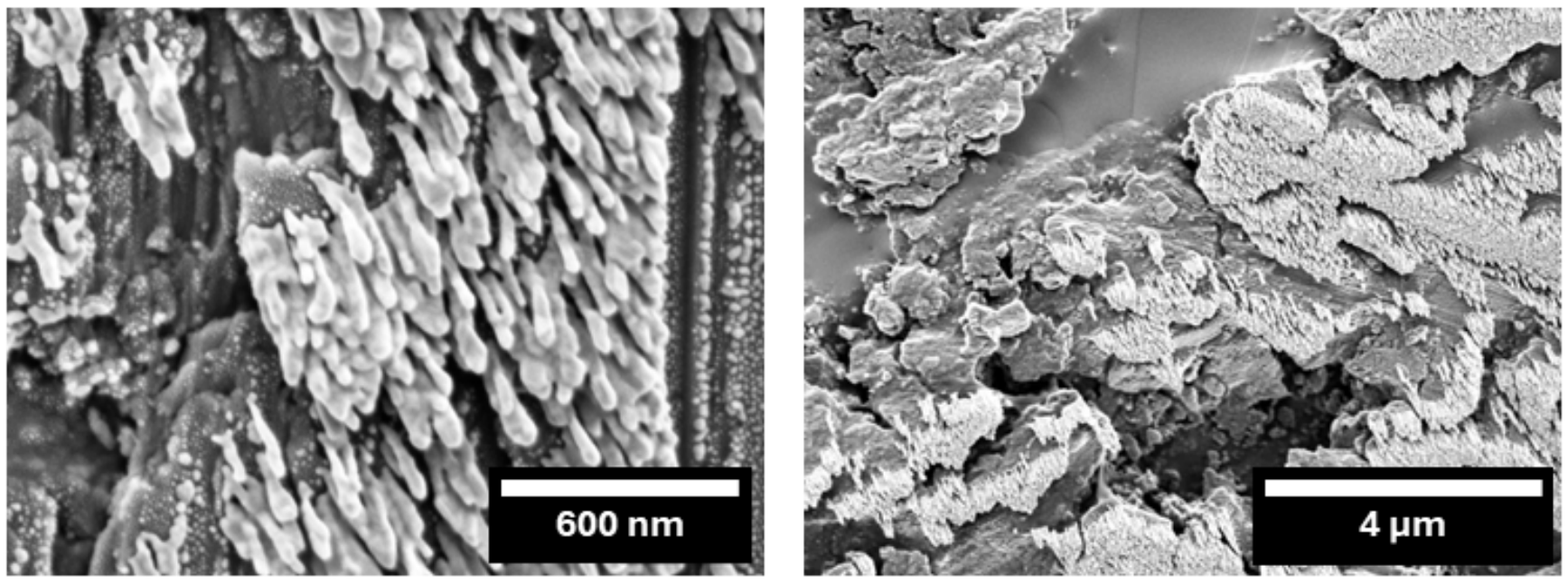

**Figure S3.** SEM image of AuNC-Ti (rough) surface at various magnifications.

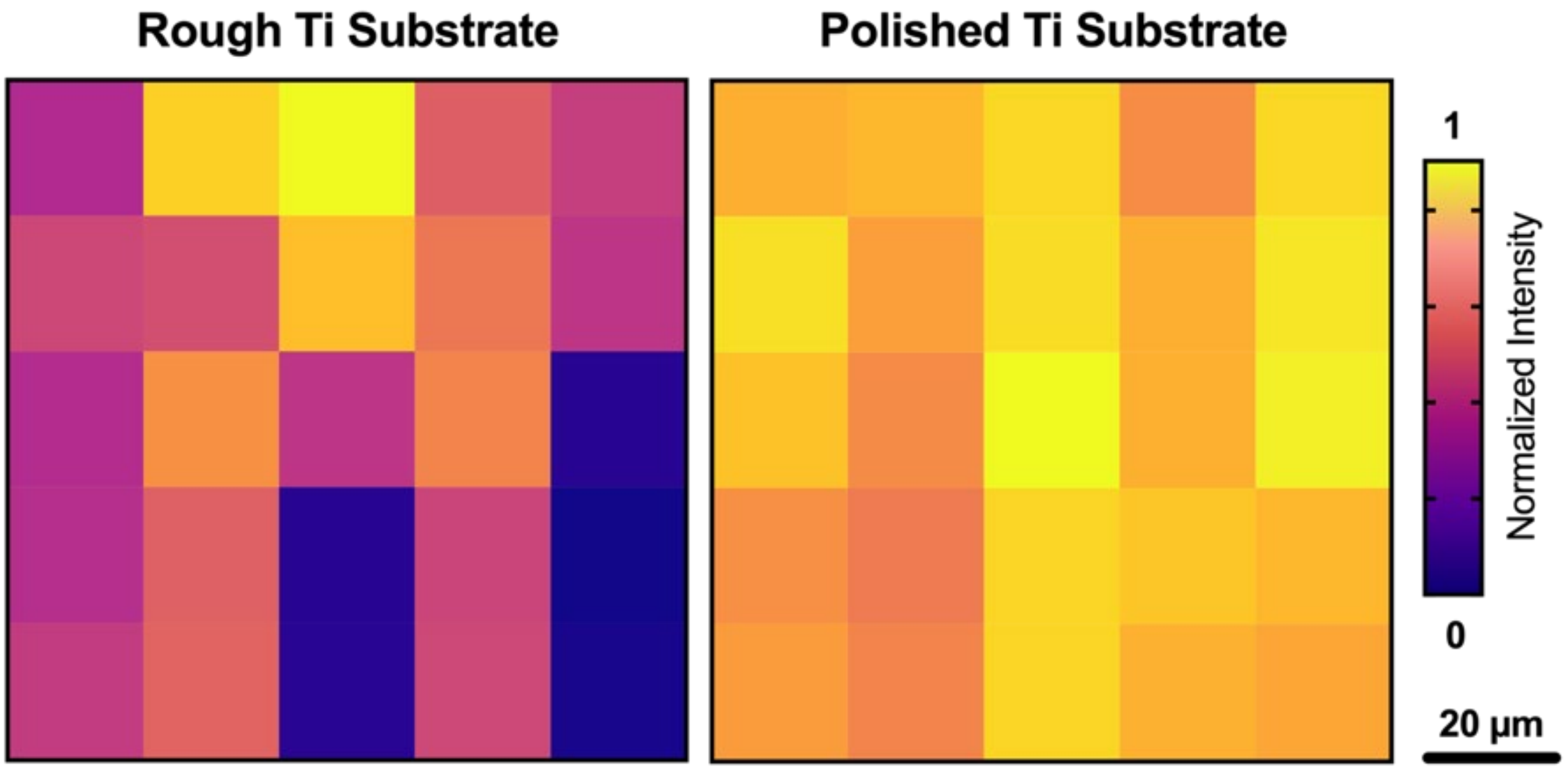

**Figure S4.** Raman maps of 4-MBA coated AuNC-Ti surface comparing spatial distribution of Raman enhancement between surfaces of AuNCs coated on a rough and polished Ti substrate. Color scale is normalized to the maximum intensity of the 1075 $cm^{-1}$ 4-MBA peak of each 100 μm by 100 μm.

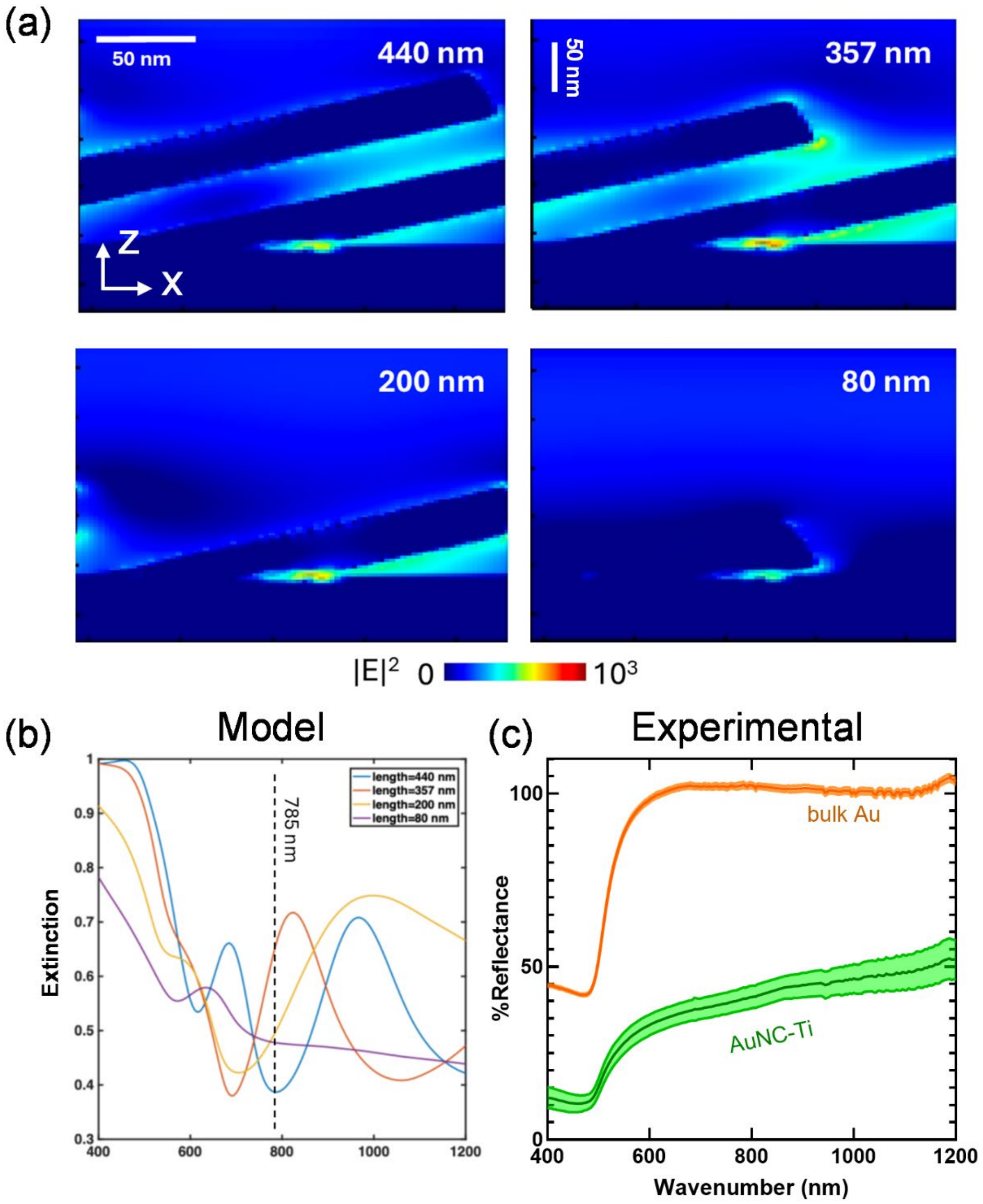

**Figure S5.** FDTD simulations comparing plasmonic and optical properties under x-polarized light-source with respect to varying AuNC lengths and experimental reflectance measurements of AuNC-Ti surface. **a)** Electric field intensity maps under 785 nm excitation and **b)** extinction spectra of AuNC-Ti using finite-difference time-domain method (FDTD) simulations. AuNCs lengths were varied to show the effect of nanoparticle aspect ratio on optical properties and electric field enhancement. AuNC width, angle, and surface density were modeled after average experimentally measured values (50 nm width, 22° tilt, and 70 AuNCs per $\mu m^2$). The AuNC length of 357 nm (representing experimental length) exhibited the highest electric field intensity and extinction within the 785 nm range, indicating it as a promising candidate design for SERS applications. **c)** Experimental reflectance spectra of smooth bulk Au as reference and AuNC-Ti using under a nonpolarized light source. Error bars represent standard deviation. Absorption was observed in NIR region partially corroborating the 357 nm AuNC length extinction model but instead exhibited a broadband response.

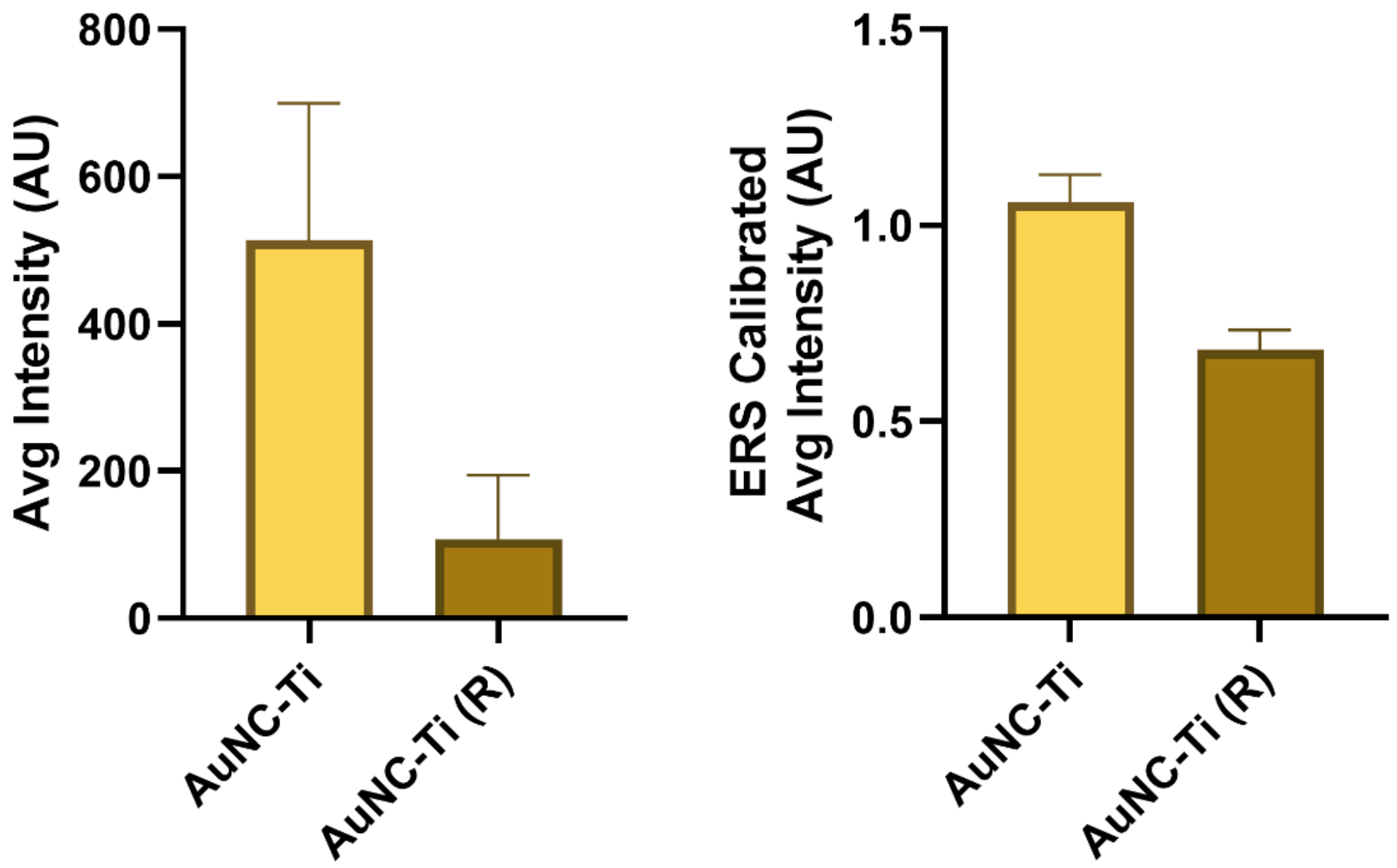

**Figure S6.** Comparison of average Raman intensities of 100 μm by 100 μm scans of 4-MBA coated polished and rough AuNC-Ti surfaces before and after calibration with plasmon-enhanced electronic Raman scattering (ERS) pseudo peaks. Raman signals were averaged from individual 25 pixel 100 μm by 100 μm area scans at 1075 $cm^{-1}$ after incubation in 4-MBA for 1 hr. ERS pseudopeaks at ~60 $cm^{-1}$ were used for calibration in accordance to Nam et al's method.[1] N = 9 scans were performed per group.

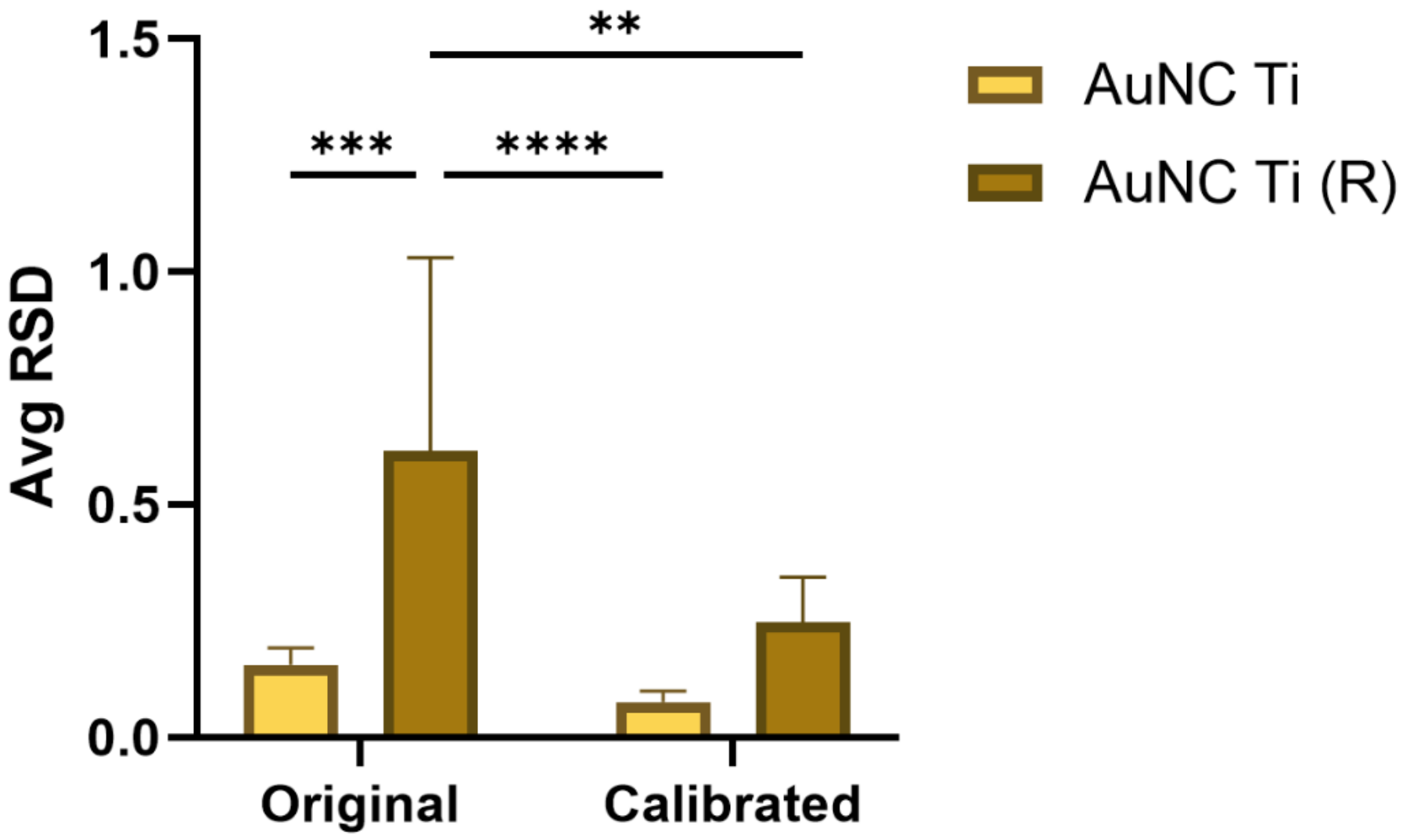

**Figure S7.** Comparison of relative standard deviations (RSD) of Raman enhancement of 4-MBA monolayer within an area scan on polished and rough AuNC-Ti surfaces before and after calibration with plasmon-enhanced electronic Raman scattering (ERS) signals.[1] RSDs were obtained from the 1075 $cm^{-1}$ peak after coating in 4-MBA, scanning a 25 pixel 100 µm by 100 µm area, and calculating the percent standard deviation within this area. N = 9 scans were performed per group.

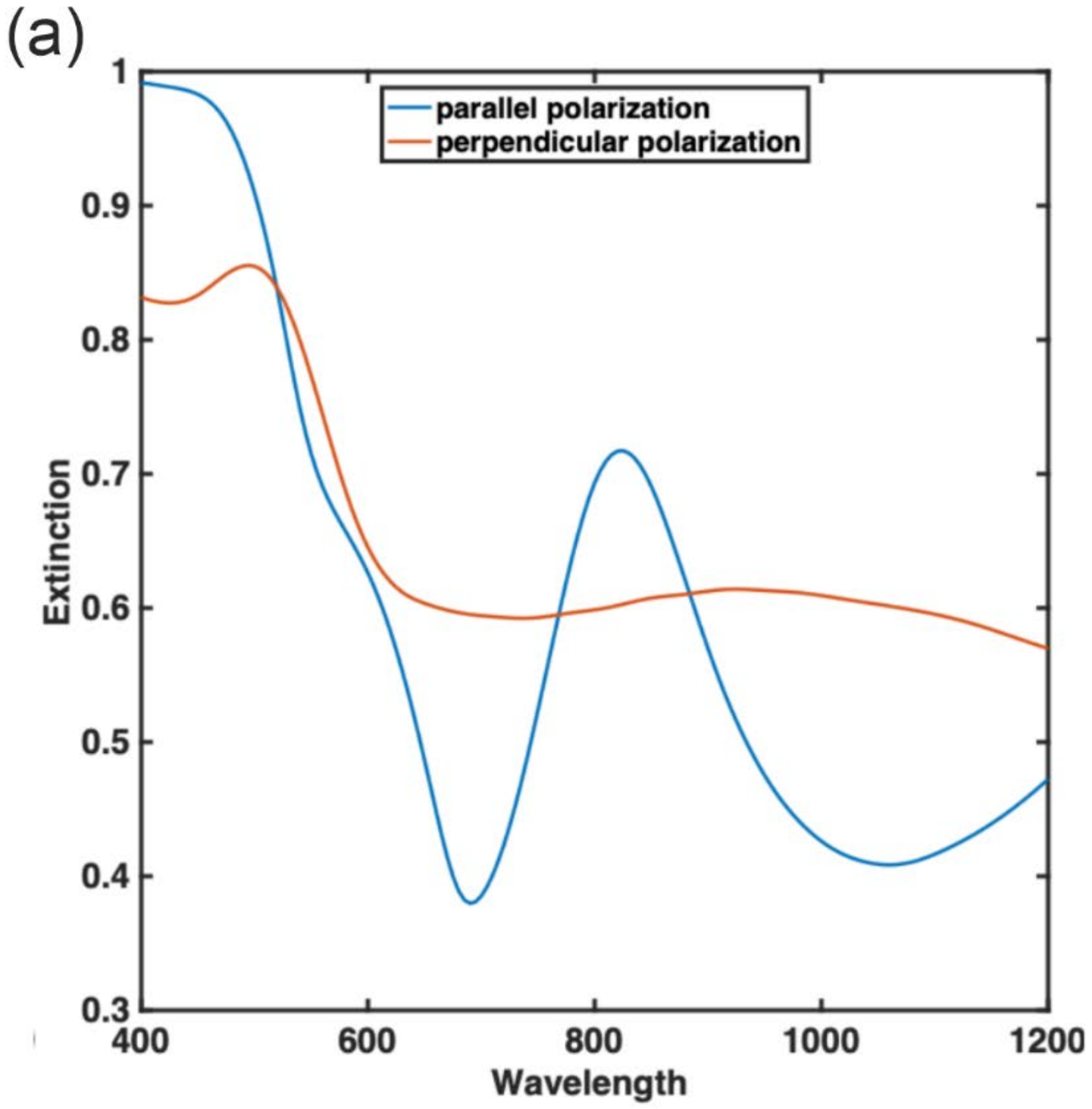

**Figure S8.** Modeled optical and plasmonic responses of AuNC-Ti material with respect to normal excitation of x- and y-polarized light. FDTD simulated **a)** extinction spectra and **b)** electric field intensity maps of AuNC-Ti surface excited by a polarized light source perpendicular and parallel to AuNC orientation. Intensity maps were generated under 785 nm excitation. Parallel polarization coupled strongly with longitudinal plasmonic modes compared with perpendicular polarization.

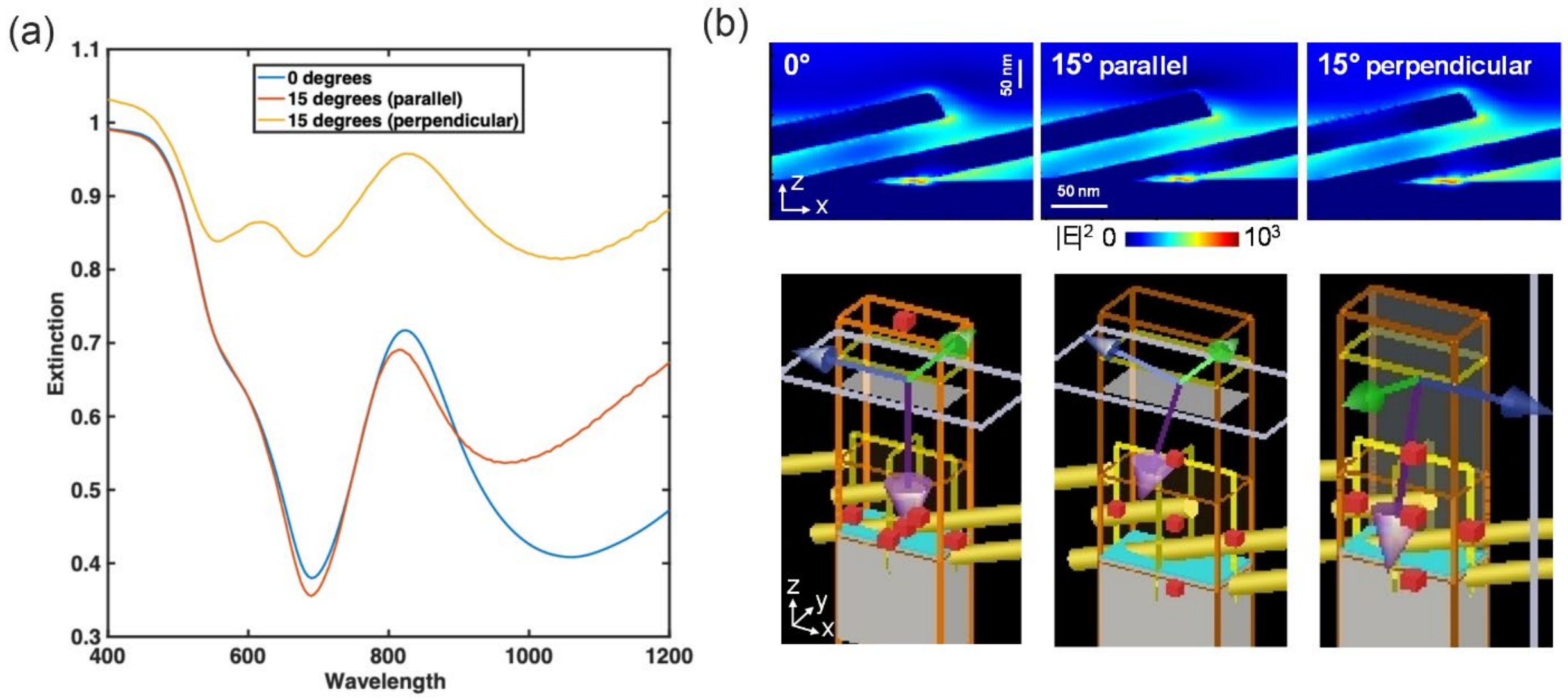

**Figure S9.** FDTD simulated plasmonic and optical response of AuNC-Ti under various incidence beam orientations. **a)** Extinction spectra and **b)** electric field intensity maps of AuNC-Ti surface excited by s-polarized wave source at 0° and 15° incident angle. Intensity maps were generated under 785 nm excitation.

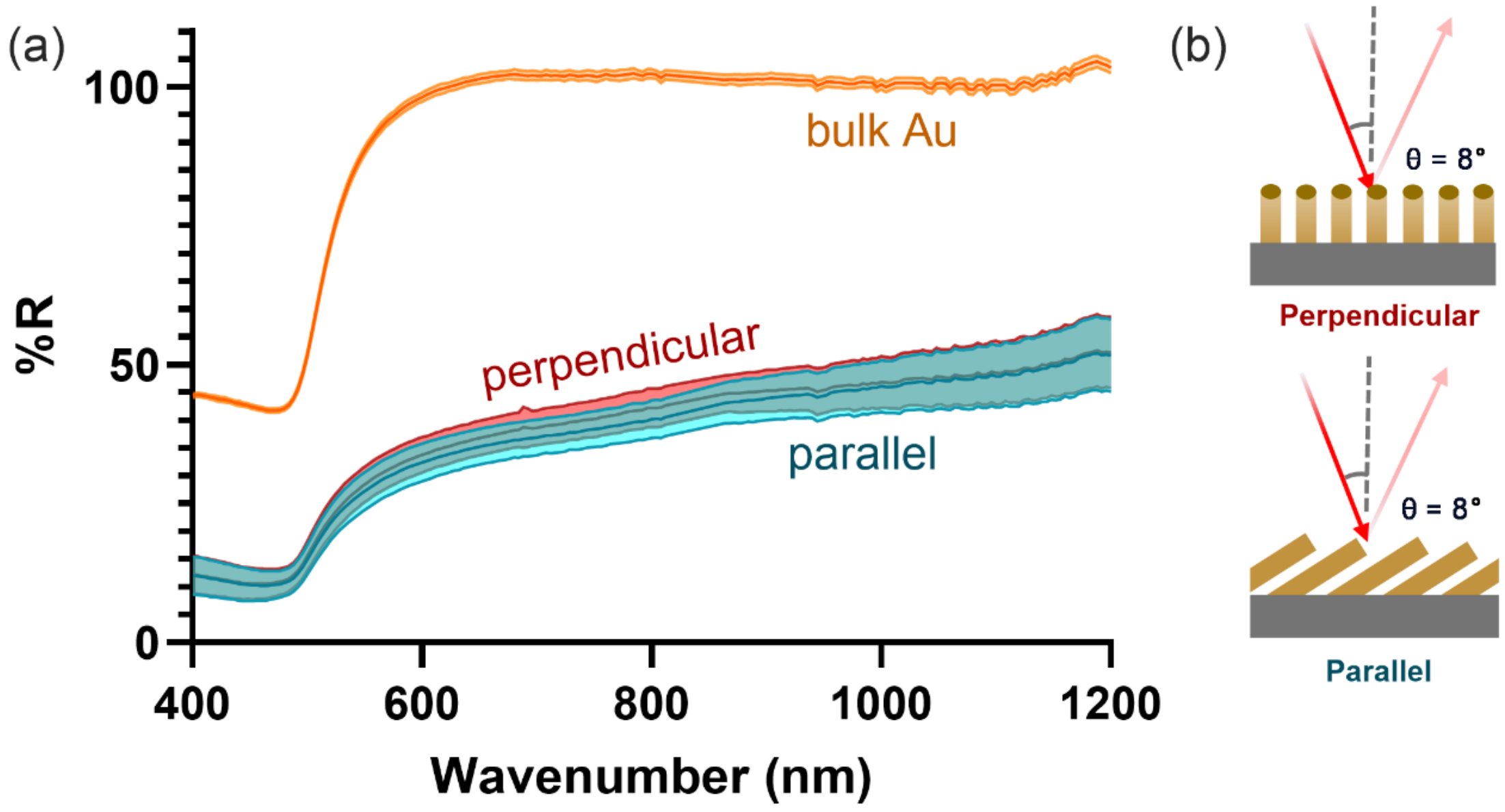

**Figure S10.** Reflectance spectra at perpendicular and parallel incidence beam orientations with respect to AuNC-Ti direction under nonpolarized light source. **a)** 8° incident angle reflectance measurements taken of AuNC-Ti surface under perpendicular and parallel orientations and compared with smooth gold reflectance measurements. Error bars represent standard deviation. **b)** Illustration of respective parallel and perpendicular incident beam orientations to with respect to AuNC direction performed during reflectance measurements.

**Table S1.** XPS elemental percent composition of surfaces, including rough Ti substrates.

| | C1 s | O1 s | Ti 2p | Au 4f |
|---|---|---|---|---|
| **Bare Ti** | 19.2 ± 0.2 | 60.1 ± 0.4 | 20.8 ± 0.3 | N/A |
| **AuNC-Ti** | 32 ± 4 | 10 ± 3 | 1.1 ± .9 | 57 ± 6 |
| **Bare Ti (R)** | 32 ± 3 | 47 ± 2 | 21.4 ± .7 | N/A |
| **AuNC-Ti (R)** | 32 ± 1 | 20 ± 2 | 6.3 ± .6 | 42 ± 2 |

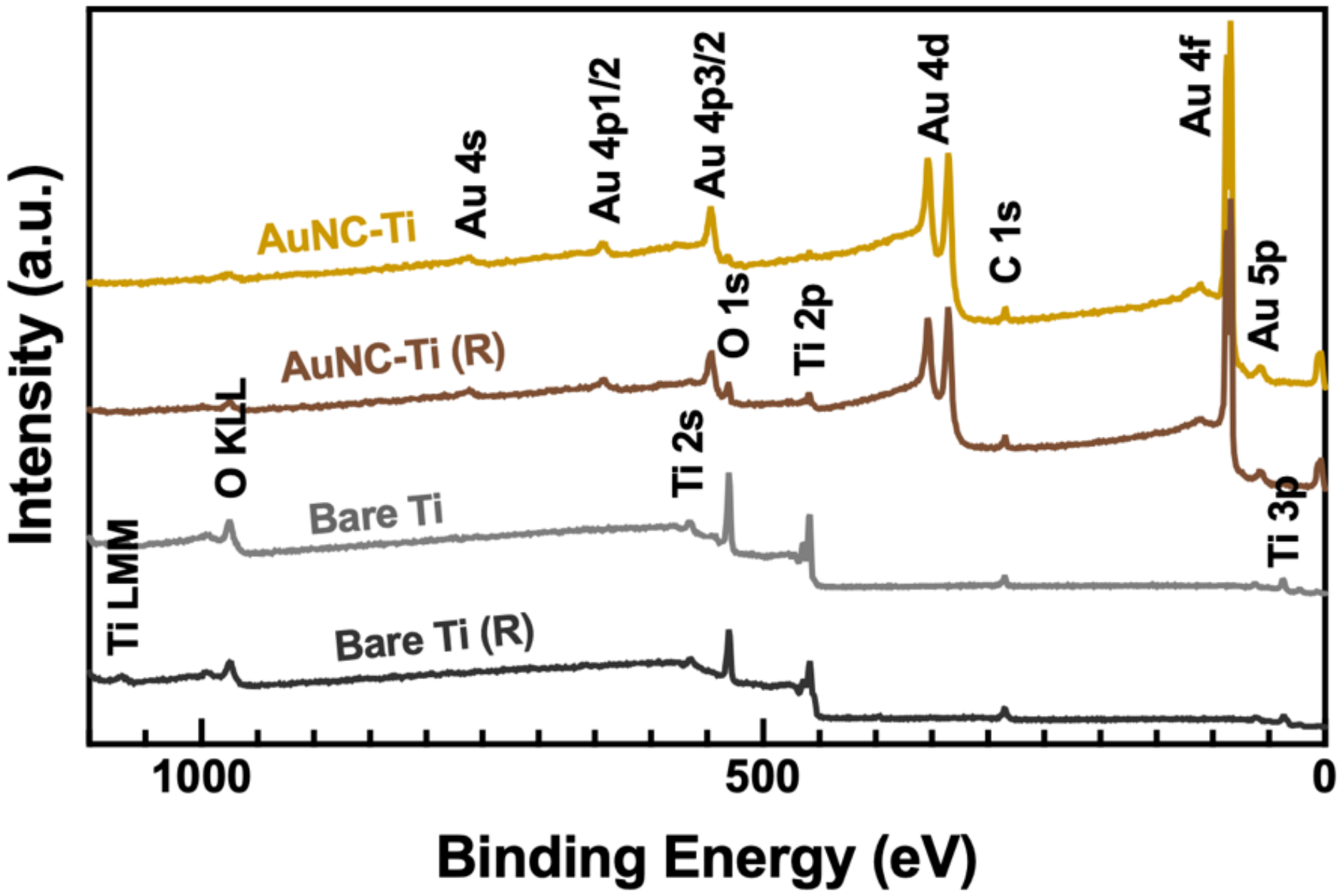

**Figure S11.** XPS survey scans over AuNC-Ti surfaces and bare Ti surfaces with a polished Ti substrate (AuNC-Ti) and rough Ti substrate (AuNC-Ti (R)).

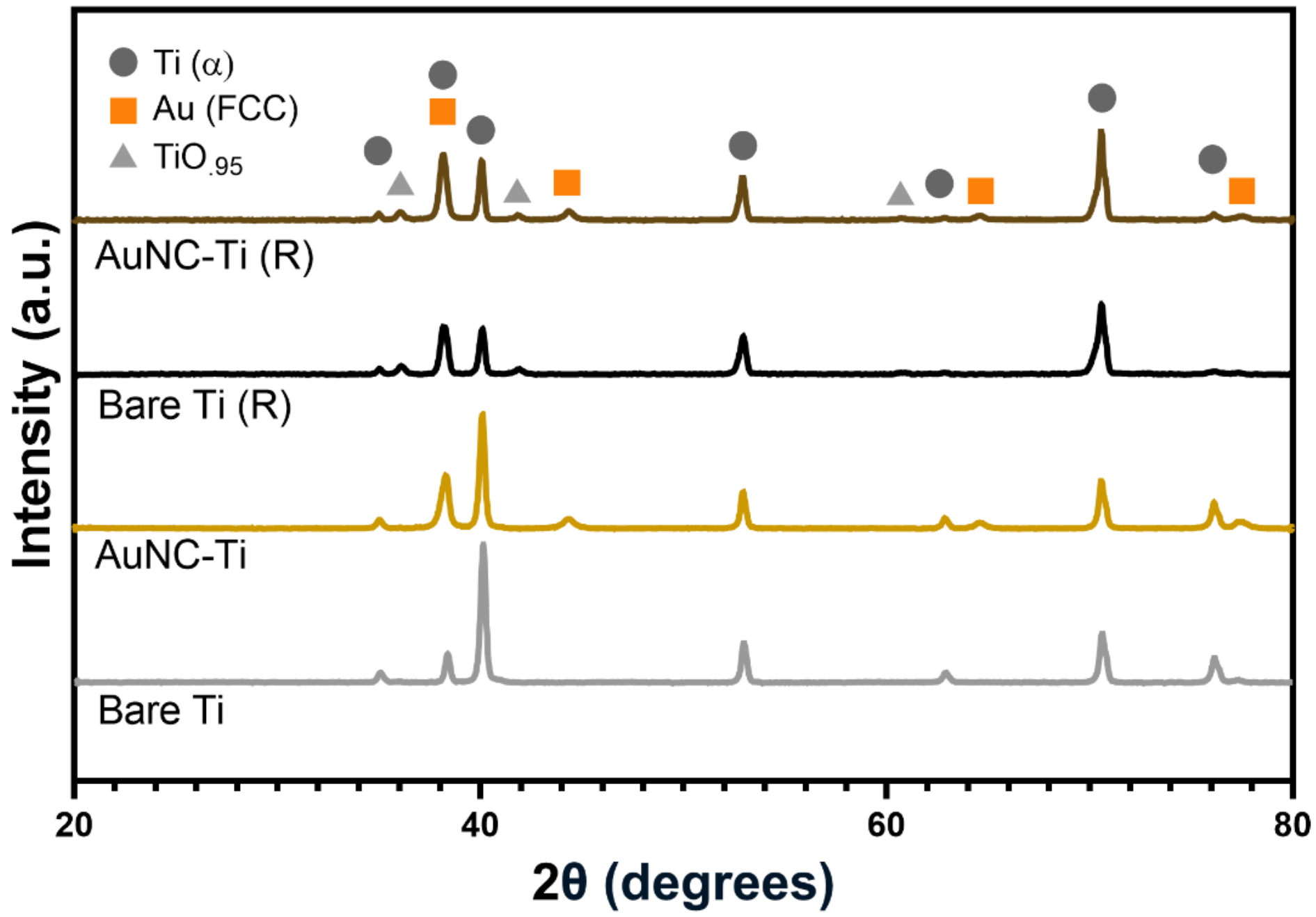

**Figure S12.** XRD patterns of AuNC-Ti and bare Ti surfaces with and without polishing of Ti substrate.

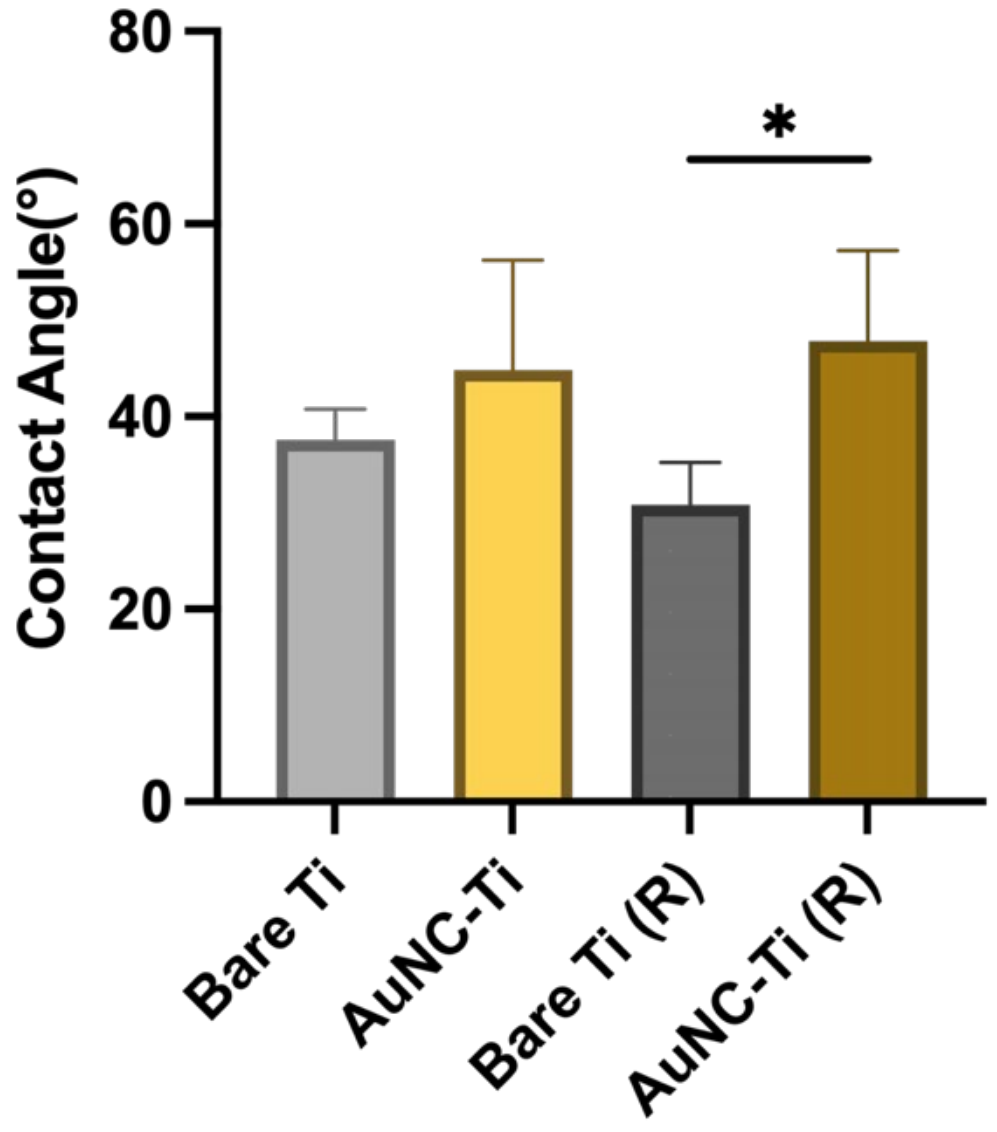

**Figure S13.** Contact angle measurements for bare Ti and AuNC-Ti surfaces with and without Ti polishing.

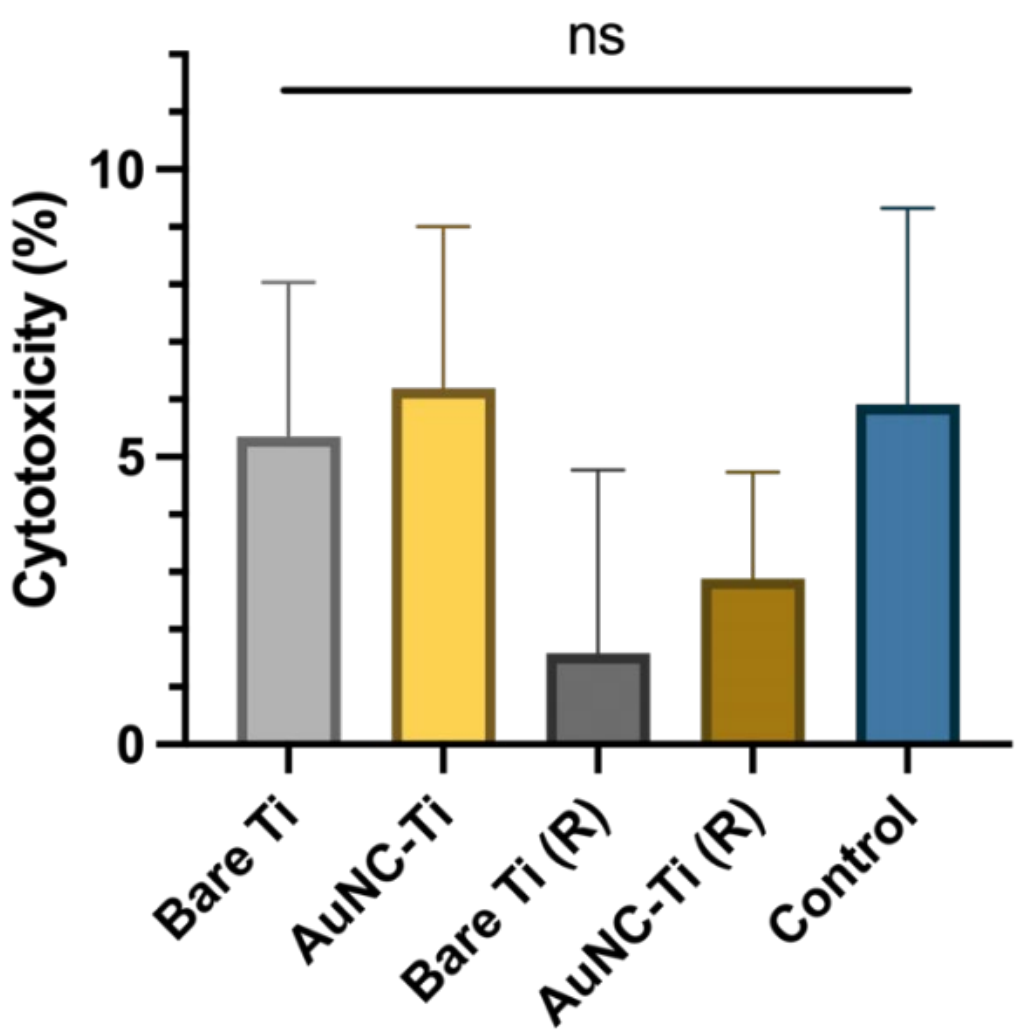

**Figure S14.** Percent cytotoxicity of bare and AuNC-coated Ti with and without Ti polishing, compared to cell and media only control.

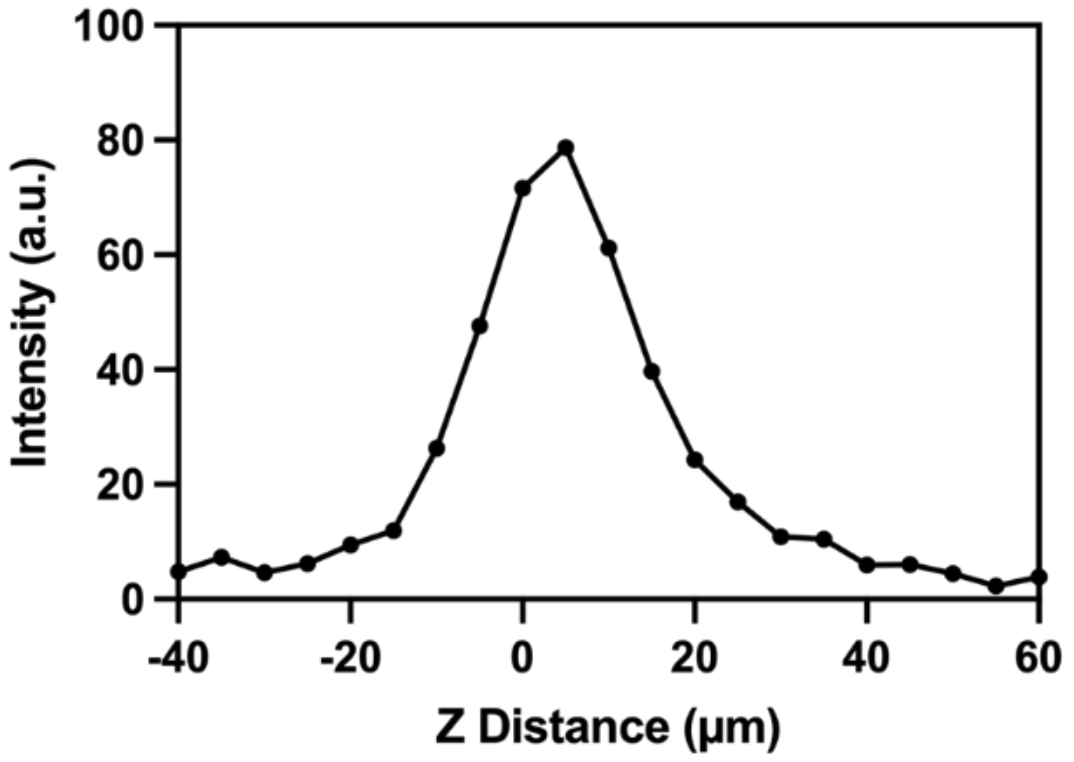

**Figure S15.** Measured Raman intensity of Si at 527 $cm^{-1}$ acquired over a z-scan on the bare silicon wafer surface. The z-scan was performed in 5 μm increments over 100 μm, and each measurement was averaged over 20 accumulations. A 785 nm lazer at 5.74 mW power, 300 g/mm grating, and 20x magnification were used. The effective scattering volume height ($H_{eff}$) was found by dividing the area under the curve (total intensity of effective volume) by the maximum intensity.[2,3] $H_{eff}$ was calculated to be 28.6 μm.

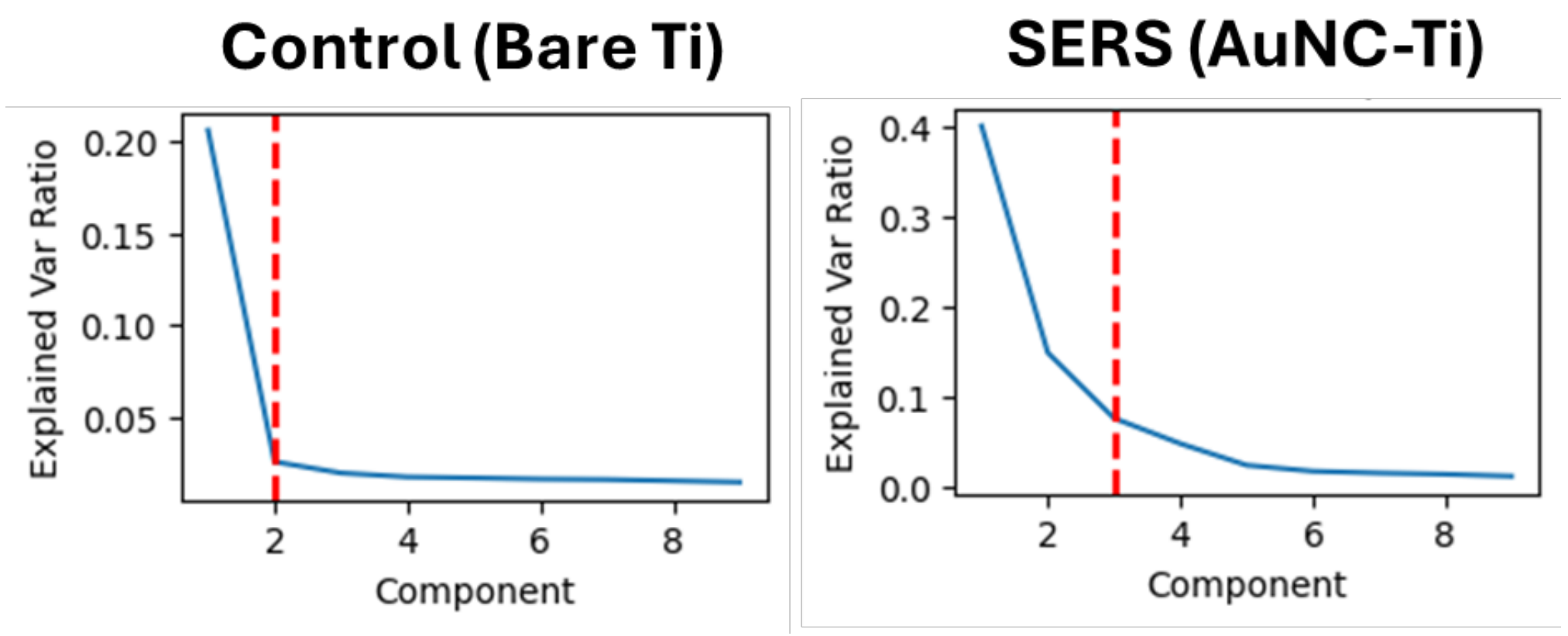

**Figure S16.** Elbow plots of the first nine principal components (PCs) of Raman maps taken of tissue on bare Ti and AuNC-Ti over 400 pixels with a 300 μm by 300 μm area. As an empirical method, the elbow method is effective when there is a clear bend or "elbow" when plotting explained variance against individual components in order of decreasing explained variances. While the elbow point was visually apparent in the majority of the data, a heuristic to identify the elbow point consistently in each map was used. The point where the change in explained variance between consecutive components (or the slope) was less than 10% of the maximum explained variance (in PC1) was considered the elbow point.

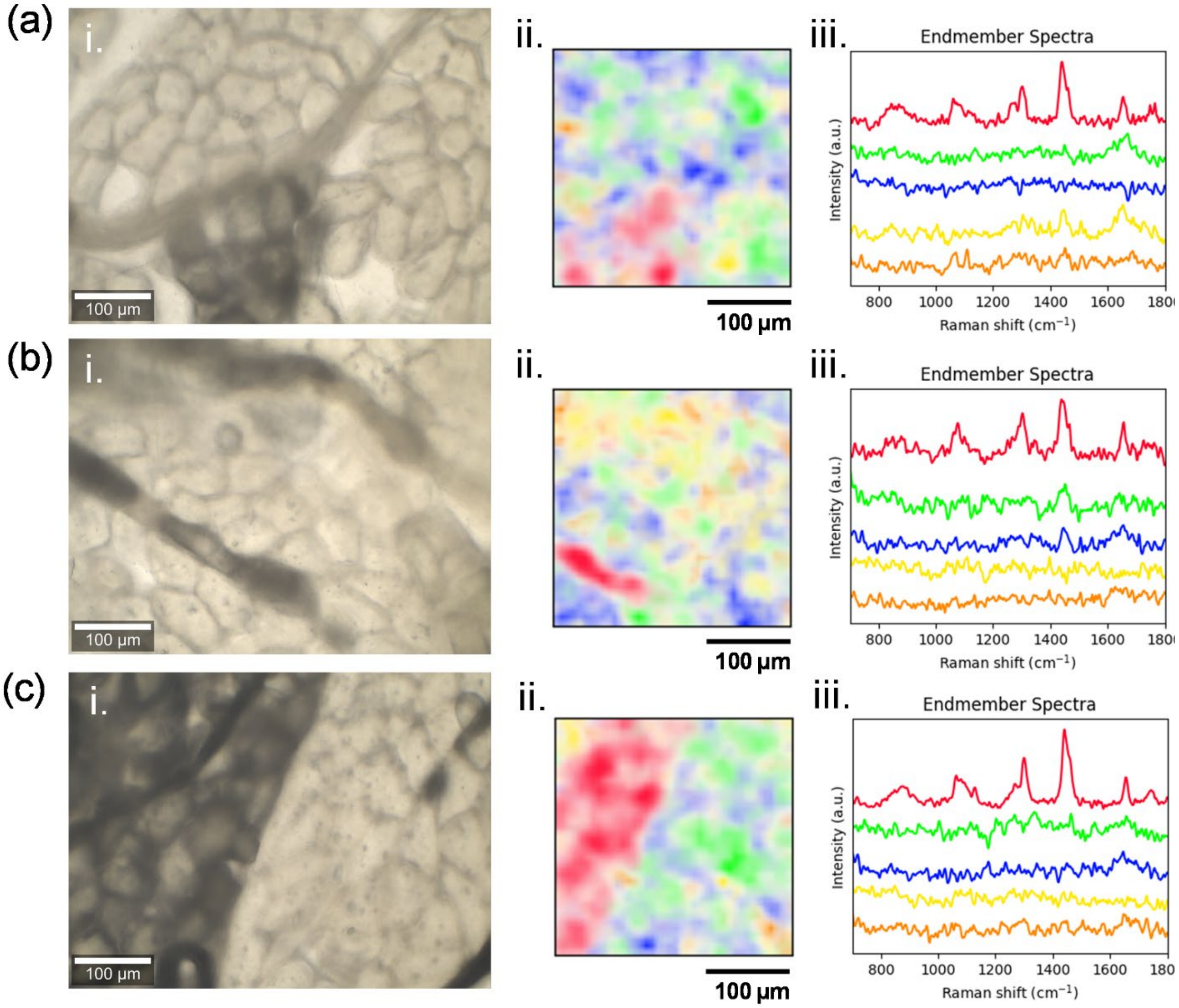

**Figure S17.** i) Additional sample images and ii) corresponding abundance maps and iii) endmember spectra for pork shoulder tissue placed on the control bare Ti surface **(a, b, and c denote individual scan areas).**

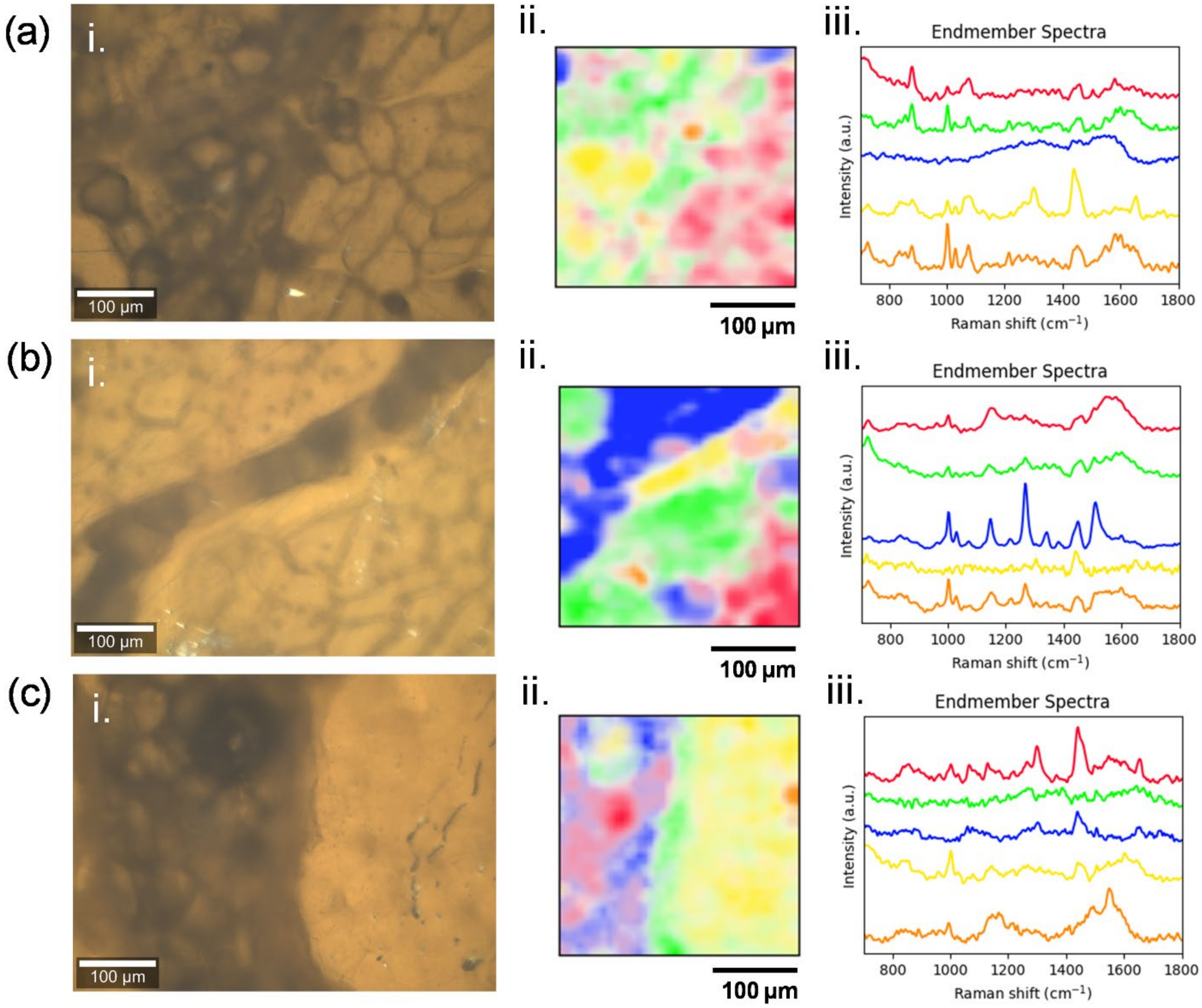

**Figure S18.** i) Additional sample images and ii) corresponding abundance maps and iii) endmember spectra for pork shoulder tissue placed on the AuNC-Ti surface **(a, b, and c denote individual scan areas).**

## SUPPLEMENTARY DISCUSSION

### *Rough vs Polished Ti substrate comparison*

Rough, unpolished Ti substrates were of interest to us for their microscale topography. Surface roughness on both the micro and nano scale improves cellular adhesion and promotes osseointegration, so we additionally characterized AuNC-Ti without the initial Ti polishing step (AuNC-Ti (R) to denote AuNC-coated rough Ti substrate).[4] SEM images of these Ti substrates are shown in **Figure S2** to illustrate the difference surface roughness. AuNC-Ti (R) surfaces were compared with AuNC-Ti using polished Ti (denoted as just AuNC-Ti). The main observations include greater stability of AuNC-Ti (R) materials, but lesser intensity and greater heterogeneity of SERS signals. ERS normalization was able to improve uniformity of SERS enhancement (**Figure S6**), but the lower intensity of Raman signals from the AuNC-Ti (R) material was a concern for in-vivo use, where SERS signals must overcome strong surface scattering. XPS and XRD data (**Figure S11 and S12**) indicate a higher atomic surface concentration of oxide and Ti on the AuNC-Ti (R) surface compared to the smooth AuNC-Ti surface. This was likely due to poorer AuNC coverage on the rough substrate, exposing the Ti and oxide layer below. $TiO_{.95}$ related peaks were also found in XRD spectra collected from unpolished surfaces. $TiO_{.95}$ is typically formed at 1500°C from $TiO_2$, so the presence of this layer was likely due to the manufacturing process from the supplier which was removed during polishing.[5] Hydrophilicity and cytotoxicity of the AuNC-Ti (R) surface was comparable to the AuNC-Ti material, indicating micro-rough SERS active Ti to be a viable candidate for self-sensing implant materials. Methods to strengthen SERS enhancement of heterogeneous SERS coatings should be explored before proceeding to *in vivo* studies, however.

*Assessment of AuNC length on SERS enhancement and optical properties*

SERS EF measurements comparable within literature values of other SERS substrates suggest that the high aspect ratio of the AuNCs, which imparts biocompatibility *in vivo*, can additionally provide considerable Raman resonance under photo illumination in the near-infrared region (NIR).[6] Gold nanorods (AuNRs) of various aspect ratios (analogous to that of AuNCs) have been of great interest for their LSPR modes along the transverse and longitudinal axes.[6] The longitudinal LSPR mode wavelength in particular can be tuned from the visible to NIR region by modifying AuNR aspect ratios or lengths as explained in Mie theory.[6,7] NIR wavelengths penetrate deeper into biological tissue, making AuNRs engineered for NIR excitation of great interest for noninvasive *in vivo* SERS applications.[8]

We performed finite-difference time-domain (FDTD) simulations using models of AuNCs of various lengths on Ti with its native oxide to investigate additional shape aspect ratios potentially better suited for deep tissue imaging by computationally simulating the plasmonic effect of different lengths. Electric field intensity maps and extinction spectra of AuNCs of various lengths were simulated under 785 nm to estimate LSPR bands (**Figure S5 (a and b)**). Extinction spectra demonstrated that as AuNCs lengthened, their LSPR bands blue-shifted and narrowed. A clear transverse and longitudinal mode were observed in the 440 nm AuNC model with separate modes becoming less apparent as AuNC length decreases, which is likely due to the transverse mode blue-shifting toward the absorption band of gold around 500 nm. The longitudinal mode for the AuNC length of 357 nm (corresponding to the experimentally measured length) aligns the best with the 785 nm range, suggesting that this length could be optimal for NIR excitement. The slight red shift in absorption from the laser source is also beneficial for enhancing Raman emission wavelengths in addition to excitation. Electric field maps agreed with the extinction spectra, as the 357 nm AuNC model exhibited the highest electric field intensity under 785 nm illumination.

Experimental reflectance measurements were performed on our AuNC-Ti material using UV-Vis-NIR spectrometry to compare with the modeled extinction spectra to verify the computational model. Experimental spectra exhibited a broad band response, unlike the predicted model spectra, likely due to the heterogeneity in AuNC sizes and inter-nanoparticle distances. However, the experimental broadband response still overlapped with the NIR range, which implies its ability to excite under a 785 nm laser and emit relevant Raman wavelengths, which was observed during SERS measurements (**Figure 3**).

The effect of AuNP shape and size on SERS enhancement is a crucial consideration in the context of application to biomedical implants. Therefore, it is promising that the AuNC shape is both biocompatible and appropriately SERS active in the NIR range. To preserve this morphology in vivo, ultra-thin oxide coating can be employed on AuNCs to prevent detachment and also changes in size by shielding from gold-reactive ions and oxidizing agents.[9–11] In the case of slight size or shape changes, we also found that calibration using plasmon-enhanced electronic Raman scattering pseudo peaks normalizes Raman measurements with respect to changes in intrinsic plasmonic resonance imparted by the AuNCs (**Figure S6 and S7**)[1].

***Assessment of light source incidence angle on SERS enhancement and optical properties***

Incident angle of the laser during signal acquisition is an important variable to consider during *in vivo* measurements as consistent laser orientation will be difficult to achieve due to varying device surface topology and placement within the body. We investigated the effect of incident angles with respect to the anisotropic orientation of the AuNC-Ti surface on optical and plasmonic properties to understand the potential effect of incident beam orientation on SERS sensitivity. Due to physical constraints of our Raman system, laser incident angles was constricted from varying during SERS measurements. We therefore turned to computational FDTD modeling and experimental reflectance measurements to observe LSPR absorbance bands under various incidence beam orientations to estimate the effect of non-normal excitation of AuNC-Ti surfaces.

FDTD simulations were first used to model plasmonic resonance with respect to normal and nonnormal incidence angles of 15° (**Figure S9**). The model indicated that an incidence beam tilted parallel along the AuNC axis resulted in similar plasmonic absorption at 785 nm excitation compared to the model with a normal incidence angle. On the contrary, absorption when tilted perpendicular to AuNC orientation was greatly increased compared to both normal and parallel tilt orientations. This may have been due to the model's use of s-polarized light which does not necessarily recapitulate nonpolarized light sources typically used in Raman systems.

Non-normal incident beam orientations on AuNC-Ti were subsequently studied using UV-Vis-NIR spectrometry to obtain reflectance spectra. AuNC-Ti reflectance was compared with bulk smooth gold reflectance spectra as a reference (**Figure S10**). All measurements were taken at an 8° incident angle. Unpolarized light reflectance was measured with the plane of incidence perpendicular and parallel to the longitudinal axis of the AuNCs to compare plasmonic absorption. No statistically significant difference was observed between reflectance under perpendicular and parallel conditions around 785 nm, which can be expected as unpolarized light equally excites longitudinal and traverse LSPR modes assuming the beam spot size remains equal. Normal reflectance measurements could not be compared with non-normal reflectance due to limitations in the spectrometer system. Future studies involving freely oriented hand-held

Raman probes can be conducted to accurately simulate such angles to better observe changes in SERS intensity. In the case that incidence angle greatly affects SERS measurements, ERS calibration may be considered to normalize changes in plasmonic resonance due to excitation angle.

## SUPPLEMENTARY METHODS

### ***SERS Substrate Enhancement Factor (SSEF) calculations***

SERS substrate enhancement factor (SSEF) was then calculated at each location with the following equation[12,13]:

$$SSEF = \frac{I_{SERS}/N_{Surf}}{I_{RS}/N_{Vol}} = \frac{I_{SERS}/(\mu_M \mu_S A_S)}{I_{RS}/(C_{RS} H_{eff})}$$

where $I_{SERS}$ and $I_{RS}$ represent the Raman intensity of the enhanced and unenhanced analyte respectively. Intensities were measured with the characteristic peak at 1075 $cm^{-1}$ for surface enhanced 4-MBA and corresponding peak at 1097 $cm^{-1}$ in 100 mM neat 4-MBA ethanolic solution, originating from the ring-breathing modes of 4-MBA. Laser power was increased to 86.7 mW when measuring neat 4-MBA solution and background ethanol. Ethanol background was subtracted from the 4-MBA solution spectrum before determining $I_{RS}$. $I_{RS}$ and $I_{SERS}$ were normalized by laser power. 5 replicate Raman measurements were taken and averaged when measuring 4-MBA neat solution and ethanol Raman intensity. $N_{Surf}$ represents the number of absorbed analyte molecules on the SERS substrate during spectral acquisition which was approximated using $\mu_M$ (packing density for the analyte), $\mu_S$ (packing density for individual nanocolumns), and $A_S$ (scattering area of the nanocolumn structure). $N_{Vol}$ represents the number of analyte molecules within the scattering volume without enhancement in the neat solution. This was approximated using the concentration of the neat 4-MBA solution ($C_{RS}$) and the effective height of the scattering volume ($H_{eff}$) which were 100 mM and 28.6 μm respectively.[12] $\mu_M$ of 4-MBA was referenced from literature to be 5.263 $nm^{-2}$ on gold, and SEM was be used to estimate $\mu_S$ and $A_S$.[14] $A_S$ was approximated as the average surface area of the nanocolumn. $H_{eff}$ was determined by performing a z-scan and measuring the intensity of a Si wafer (527 $cm^{-1}$) which is detailed in **Figure S15**.[2,3]

*Example calculation:*

$$\mu_M = 5.263\ \mathrm{nm}^{-2} = 8.74 \times 10^{24}\ \mathrm{mol\ 4\text{-}MBA/nm}$$

$$\mu_S = 70\ \mathrm{AuNCs} / \mu\mathrm{m}^2$$

$$A_s = h_s d_s \pi + (d_s/2)^2\pi = 358\ \mathrm{nm} * 48\ \mathrm{nm} * \pi + (48\ \mathrm{nm}/2)^2\pi = 56000\ \mathrm{nm}^2$$

$$H_{eff} = 2.86 \times 10^4\ nm$$

$$C_{RS} = 100\ \mathrm{mM\ 4\text{-}MBA}$$

$$I_{RS} = 0.857\ avg\ \mathrm{CCD\ cts}$$

$$I_{SERS} = 2257\ \mathrm{CCD\ cts}$$

$$SSEF = \frac{I_{SERS}/(\mu_M \mu_S A_S)}{I_{RS}/(C_{RS} H_{eff})} = \frac{2257/(8.74 \times 10^{24}\ \mathrm{mol/nm} * 70\mu\mathrm{m}^{-2} * 56000\ \mathrm{nm}^2)}{0.857/(100\ \mathrm{mM} * 2.86 \times 10^4\ \mathrm{nm})}$$

$$SSEF = 2.2 \times 10^5$$

### *FDTD simulations and reflectance measurements*

Lumerical FDTD software (Ansys) was used for modeling electric field ($|E|^2$) intensity maps and extinction spectra of the AuNC-Ti material under an x-polarized light source parallel to the AuNC orientation. A uniform 3 nm mesh was used in x-, y-, and z-directions. The optical constants of Au were taken from Johnson and Christy.[15] All simulations used Bloch boundary condition in x and y direction with periodic lattice parameters $L_x$ = 117 nm and $L_y$ = 235 nm to approximate empirically observed surface density. A perfectly matched layer (PML) boundary condition was used in the z direction along with a polarized wave source. Reflectance measurements were taken using UV/Vis/NIR spectrometry (PerkinElmer Lambda 1050) and a 150mm InGaAs Integrating Sphere attachment (PerkinElmer). Measurements were taken at an 8° incident angle on the sample. Bulk smooth Au measurements were taken on a silicon wafer coated with 200 nm of Au using an electron-beam evaporator system (AJA International Inc.) at a normal incident bean angle. Reflectance measurements were taken using a minimum of triplicate samples per group. Comparisons between reflectances under 785 nm illumination were performed using a Welch's T-test.